\documentclass[
reprint,
superscriptaddress,
 amsmath,amssymb,
 aps,
prstab
]{revtex4-2}

\usepackage{graphicx}
\usepackage{dcolumn}
\usepackage{bm}
\usepackage[separate-uncertainty=true,list-units=repeat,multi-part-units=brackets]{siunitx}
[separate-uncertainty=true,list-units=repeat,multi-part-units=brackets]
\usepackage{booktabs}
\usepackage{tabularx}
\usepackage{tabularray}

\begin{document}

\preprint{APS/123-QED}

\title{High brightness, symmetric electron bunch generation in a plasma wakefield accelerator via a radially-polarized plasma photocathode}%

\author{J. Chappell}
\email{james.chappell@physics.ox.ac.uk}
\altaffiliation[Present address: ]
{Informatics Lab, Met Office, Exeter, Devon, United Kingdom}
\affiliation{%
 John Adams Institute for Accelerator Science and Department of Physics, University of Oxford, Denys Wilkinson Building, Keble Road, Oxford, United Kingdom
}%

\author{E. Archer}
\altaffiliation[Present address: ]
{Deutsches Elektronen-Synchrotron DESY, Notkestraße 85, 22607 Hamburg, Germany}
\affiliation{%
 John Adams Institute for Accelerator Science and Department of Physics, University of Oxford, Denys Wilkinson Building, Keble Road, Oxford, United Kingdom
}%

\author{R. Walczak}
\affiliation{%
 John Adams Institute for Accelerator Science and Department of Physics, University of Oxford, Denys Wilkinson Building, Keble Road, Oxford, United Kingdom
}%
\affiliation{Somerville College, Woodstock Road, Oxford, United Kingdom}

\author{S. M. Hooker}
\affiliation{%
 John Adams Institute for Accelerator Science and Department of Physics, University of Oxford, Denys Wilkinson Building, Keble Road, Oxford, United Kingdom
}%

\date{\today}

\begin{abstract}
The plasma photocathode has previously been proposed as a source of ultra-high-brightness electron bunches within plasma accelerators. Here, the scheme is extended by using a radially-polarized ionizing laser pulse to generate high-charge, high-brightness electron bunches with \textit{symmetric} transverse emittance. Efficient start-to-end modelling of the scheme, from ionization and trapping until drive bunch depletion, enables a multi-objective Bayesian optimisation routine to be performed to understand the performance of the radially-polarized plasma photocathode, quantify the stability of the scheme, and explore the fundamental relation between the witness bunch charge and its emittance. Comparison of plasma photocathodes driven by radially- and linearly-polarized laser pulses show that the former yields higher brightness electron bunches when operating in the optimally-loaded regime.
\end{abstract}

\maketitle


Light sources such as X-ray free electron lasers (X-FELs)~\cite{Pellegrini2016,Seddon_2017} and synchrotrons~\cite{Tetsuya2019, Shin2021} have transformed our understanding as essential tools for studies of biological, chemical and high energy density science. Generating this radiation requires the highest quality electron bunches, achieved by maximising their charge density in transverse and longitudinal phase space. The metric used to quantify bunch quality is the 6D \textit{brightness}, calculated via the equation~\cite{Habib2023AnnPhys}:

\begin{equation}
    B_\mathrm{6D} = \frac{2 I_\mathrm{pk}}{\varepsilon_{\mathrm{n},x} \cdot {\varepsilon_{\mathrm{n},y}}} \cdot \frac{1}{\sigma_{E,\mathrm{rel}} \;[0.1\%]},
\end{equation}

\noindent where $I_\mathrm{pk}$ is the peak current of the bunch, $\varepsilon_{\mathrm{n},x/y}$ is the transverse normalised emittance in the $x$/$y$ plane, and $\sigma_{E,\mathrm{rel}}$ is the relative energy spread of the bunch. State-of-the-art X-FEL facilities, powered by radio-frequency (RF) accelerating cavities, can achieve projected --- i.e. calculated over the entire bunch --- 6D brightnesses of $B_\mathrm{6D} \sim \mathcal{O}(\SI{1e16}{}\,\mathrm{A m^{-2}\, 0.1\%^{-1}})$, with peak longitudinal slice brightness exceeding $B_\mathrm{6D} \sim \mathcal{O}(\SI{1e17}{}\,\mathrm{A m^{-2}\, 0.1\%^{-1}})$~\cite{dimitri2014}. The use of RF accelerating cavities with field gradients typically in the range $10-100$\,\SI{}{MVm^{-1}} at such facilities requires extended linear accelerator lengths to achieve the multi-GeV electron energies needed to generate radiation with the $\lambda_r \sim\,$nm wavelengths of interest.

Plasma accelerators~\cite{TajimaDawsonPRL1979} offer a compelling alternative route to future-generation light sources thanks to the extreme accelerating gradients they can support ($1-100\,\mathrm{GVm^{-1}}$), enabling construction of high-quality, tunable light sources at much reduced cost and within a significantly more compact footprint compared to conventional accelerator facilities, potentially enabling wider access to such sources. Furthermore, beam-driven plasma accelerators~\cite{Ruth1984,ChenPRL1985} offer the potential to transform already-existing light sources, either by acting as a ``plasma booster" and to increase the energy reach of such facilities~\cite{Chen2001, Blumenfeld2007}, or through the generation of electron bunches with unique parameters, currently inaccessible to conventional RF accelerator technology. 

One such example of the latter case is the plasma photocathode~\cite{Hidding2012PRL}, a scheme in which an intense, ultra-short laser pulse co-propagates within the wakefield driven by a high-density electron bunch. Around its focus, the laser pulse selectively ionizes a high intensity threshold (HIT) dopant gas and releases electrons into the wake which can then be trapped and accelerated. A key advantage of this scheme compared to traditional photoinjectors in RF accelerators is that the electrons are released directly into the extreme longitudinal fields provided by the plasma wakefield, providing rapid acceleration and minimising space-charge effects that cause significant emittance growth. Additionally, tuning the laser pulse intensity to be just above the ionization threshold of the HIT dopant releases electrons within a small volume and can minimise their thermal emittance~\cite{Schroeder2014}. Previous simulation studies~\cite{Hidding2014, Habib2023, Habib2023AnnPhys} of the plasma photocathode scheme have indicated 6D brightness values in excess of $\SI{1e18}{}\,\mathrm{A m^{-2}\, 0.1\%^{-1}}$, a significant performance increase over the bunches provided by state-of-the-art RF facilities, in only a fraction of the footprint.

However, the generation of electron beams with such exceptional quality via the plasma photocathode scheme requires low bunch charge ($\lesssim \SI{1}{pC}$) to minimise the transverse emittance and generate such extreme brightness. This leads to significant growth of the energy spread of the electron bunch during acceleration over extended distances due to the variation of the longitudinal field amplitude over the length of the bunch itself. This makes transport of the bunch outside of the plasma accelerator particularly challenging due to the inherent chromaticity of traditional transport optics, leading to emittance growth and decreased bunch brightness. An extension of the original plasma photocathode scheme~\cite{Habib2023} proposed injection of a second, higher charge and lower quality bunch to ``escort" the high-quality witness bunch. The current of the escort bunch located at the rear of the wakefield would perturb the trajectories of in-flowing plasma electrons, modifying its structure via beam-loading~\cite{Tzoufras2008}. By appropriately tailoring the charge distribution of the escort bunch, it was shown that it is possible to minimise the variation of the longitudinal wakefield amplitude in the region of the high-quality witness bunch, maintaining its energy spread throughout acceleration. However, this scheme involves significantly increased experimental complexity. 

An alternative approach would be to increase the current of the original witness bunch such that it is capable of optimally-loading the longitudinal wakefield itself. This would require increasing the amount of charge released into the wakefield by the ionizing laser pulse by increasing its intensity, which can have a detrimental effect on the quality of the electron bunch. Owing to conservation of transverse canonical momentum, electrons released via field ionization gain a transverse momentum equal to the laser vector potential at the moment of ionization --- momentum that is directed in the plane of polarization. A linearly-polarized plasma photocathode therefore creates electron bunches with larger emittance in the polarization plane --- an effect that grows with increasing intensity. Such an emittance asymmetry can lead to emittance mixing over extended propagation distances~\cite{diederichs2024}, an effect particularly prevalent for high-charge, low-emittance electron bunches, the extreme space-charge fields of which can promote strong ion motion and generate non-linearly coupled transverse wakefields. This can lead to degradation of the electron bunch quality, decreasing its brightness. 

In this paper, we overcome this limitation through use of a radially-polarized plasma photocathode. For this polarization, the laser field in a given transverse plane is always parallel (or anti-parallel) to the radial vector, which causes electrons produced via laser ionization to have an azimuthally-symmetric emittance. In contrast, for linearly- or circularly-polarized pulses, the instantaneous laser field is uni-directional in a given transverse plane, and hence is not azimuthally-symmetric. Although use of a radially-polarized plasma photocathode leads to larger emittance at the point of ionization, we show that in the high-charge regime --- where the final emittance of the witness bunch is dominated by emittance growth during matching of the bunch to the transverse wakefield --- the average transverse emittance is minimised, and hence 6D brightness maximised, by using a radially-polarized ionizing laser pulse. This allows high-charge, low energy-spread operation and maximises brightness within a single injection event. Furthermore, through use of a multi-objective Bayesian optimisation routine we explore the fundamental relationship between the bunch quality and charge in the plasma photocathode scheme, study the sensitivity of the optimally-loaded regime to relative timing fluctuations, and discuss its prospects as an injector for future plasma-based accelerator facilities.

\section*{\label{sec:RPPC}The radially-polarized plasma photocathode}

The concept of the radially-polarized plasma photocathode matches that of a linearly-polarized scheme, except for the properties of the ionizing laser pulse. In this study, a radially-polarized vortex-based Laguerre-Gaussian laser pulse~\cite{Jolly2022OptLett} (denoted RPLP) is generated in simulation using the \url{LASY}~\cite{LASY} library. The pulse is constructed via the superposition of right- and left-handed circularly polarized vortex beams with Laguerre-Gaussian (LG) profiles of azimuthal order $l = \pm 1$ and radial order $n=0$. The full-width at half-maximum duration of the ionizing laser pulse was set to \SI{30}{fs}, while its central wavelength was $\lambda = \SI{800}{nm}$. For this proof-of-concept simulation the spot-size of the RPLP was $w_0 = \SI{4.5}{\micro m}$, giving a Rayleigh range of $z_\mathrm{R} = \SI{80}{\micro m}$, and the pulse contained a total energy of $E_\mathrm{Laser} = \SI{0.5}{mJ}$. The injection process, which occurrs within the first \SI{1}{mm} of propagation of the drive bunch, is simulated using \url{FBPIC}~\cite{FBPICLehe2016, FBPICJalasPoP2017, FBPICKirchen2020PRE}, a quasi-3D particle-in-cell (PIC) code, using $N_m = 3$ azimuthal modes. After extraction of the resulting drive and witness bunches from this simulation, their subsequent evolution is then simulated using \url{Wake-T}~\cite{Wake-TFerranPousa2019} until the point of drive bunch depletion ($z = \SI{90}{mm}$). The RPLP is not included in the \url{Wake-T} simulation since it diffracts sufficiently by $z= \SI{1}{mm}$ ($\sim 10 z_\mathrm{R}$) that its fields no longer ionize the dopant nor modify the wakefield.

Within this study, pre-ionized hydrogen of on-axis density $n_0 = \SI{1e17}{cm^{-3}}$ is assumed. This gives a wakefield cavity length, $\lambda_{p0} = 2 \pi c / \omega_{p0} = \SI{105}{\micro m}$ --- where $\omega_{p0} = \sqrt{(n_0 e^2)/(m_e \varepsilon_0})$ --- that matches the minimum electron bunch lengths available at current beam-driven plasma accelerator facilities~\cite{FLASHForwardDArcy2019,FACET-II}, and thus reduces the computational cost of the simulation. We note that the plasma photocathode scheme can be used over a range of plasma densities, and the operating density is often chosen according to the drive electron bunch parameters. Higher density operation ($n_0 \sim \SI{1e18}{cm^{-3}}$) is made possible through the use of ultra-short (few fs) drive bunches generated in a laser-wakefield accelerator --- akin to the ``hybrid" plasma accelerator scheme currently being explored~\cite{Kurz2021,Foerster2022} --- and gives access to the highest amplitude accelerating gradients as the maximum field supported by a cold plasma scales as $E_{0} \propto \sqrt{n_0}$. Lower density operation ($n_0 \sim 10^{15} - 10^{16}$\,cm$^{-3}$) is also possible using longer drive bunches (few hundred fs) available from RF accelerator facilities and is often preferred for future linear collider designs~\cite{HALHFFoster2023} due to improved drive bunch stability, while still maintaining accelerating gradients exceeding \SI{1}{GVm^{-1}}. 

\begin{figure}[t]
	\centering
	\includegraphics[width = \linewidth]{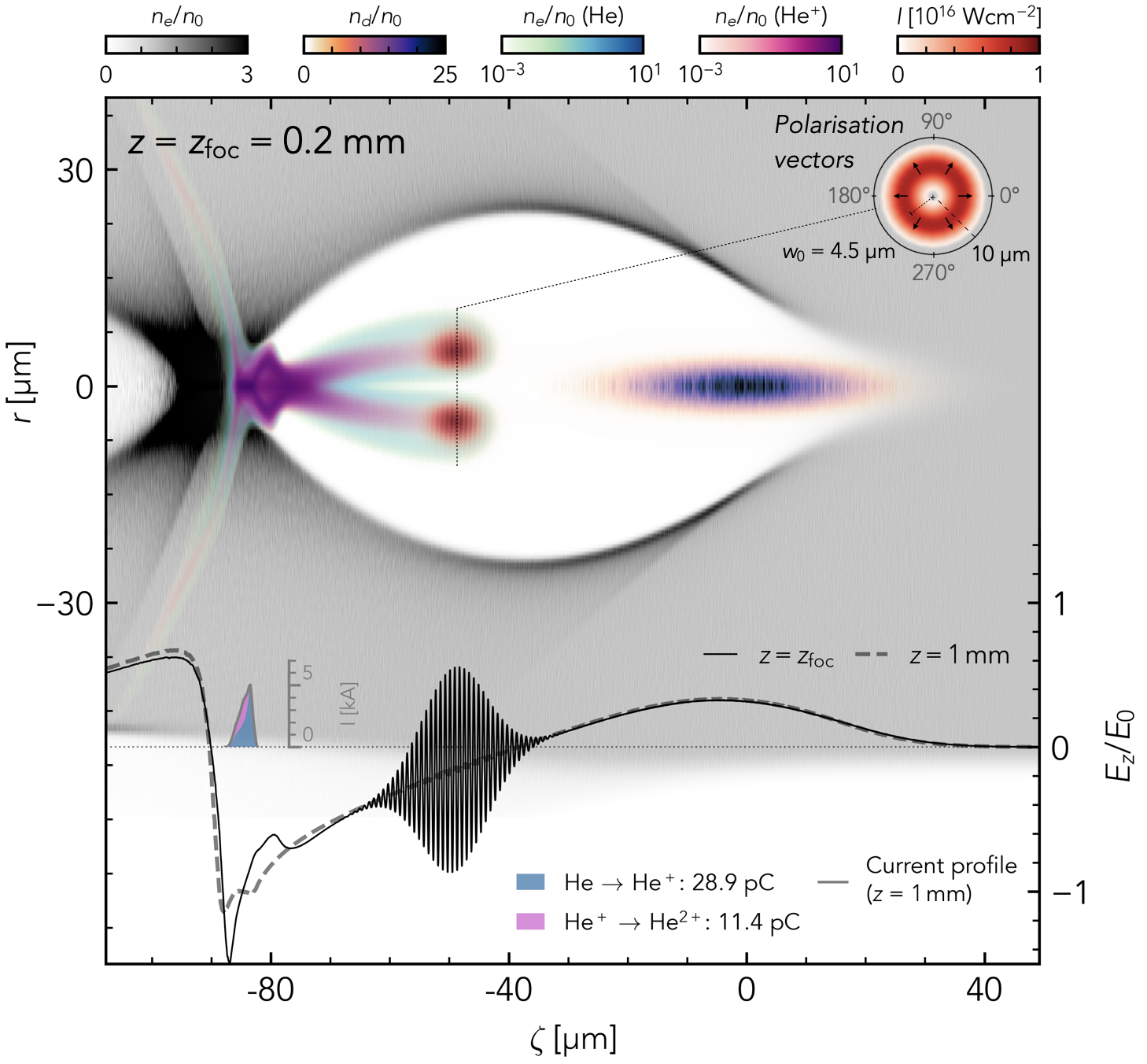}
	\caption{Snapshot from PIC simulation demonstrating the injection of dopant electrons into the wakefield driven by a high-density, ultra-relativistic electron bunch. This timestep corresponds to the focal location of the RPLP ($z = z_\mathrm{foc}$). Upper inset: The transverse intensity profile of the plasma photocathode at focus. Lower inset: The on-axis longitudinal electric field at $z = z_\mathrm{foc}$ (black) and $z = \SI{1}{mm}$ (grey, dashed), including the current profile of the trapped charge ($z = \SI{1}{mm}$). The electron number density, $n_e$, is coloured according to ionization source: pre-ionized hydrogen (grey), He (green/blue) and He$^+$ (pink/purple) and normalized to the background plasma density, $n_0 = \SI{1e17}{cm^{-3}}$. The normalized drive bunch density, $n_d$, is represented by the orange/black colormap, while the intensity profile of the RPLP is shown in red.
 }
	\label{fig:overview}
\end{figure}

Here, a matched, high-density, ultra-relativistic ($\gamma_0 = 2055$) electron bunch drives a blown-out wakefield. The RPLP follows closely, $\Delta \zeta = \SI{-45}{\micro m}$ where $\zeta = z - ct$ is the co-moving coordinate, and locally ionizes the $5\%$ neutral helium dopant over a small region around its focal position ($z = z_\mathrm{foc} = \SI{0.2}{mm}$), releasing a significant amount of charge --- initially at rest in the lab frame --- into the plasma cavity. When the focal intensity of the RPLP is sufficiently high, electrons are ionized from both the first and second levels of the helium dopant, as shown by the blue and purple regions respectively in Fig.~\ref{fig:overview}, with a reduced ionization volume of the second level due to its increased ionization threshold. The ionization energies for the first and second levels of helium are \SI{24.6}{eV} and \SI{54.4}{eV} respectively. The electrons initially slip backwards in the co-moving frame of the wakefield, and are rapidly accelerated by the longitudinal wakefield, leading to trapping of a fraction at the rear of the accelerating cavity. 

\begin{figure}[t]
	\centering
	\includegraphics[width = \linewidth]{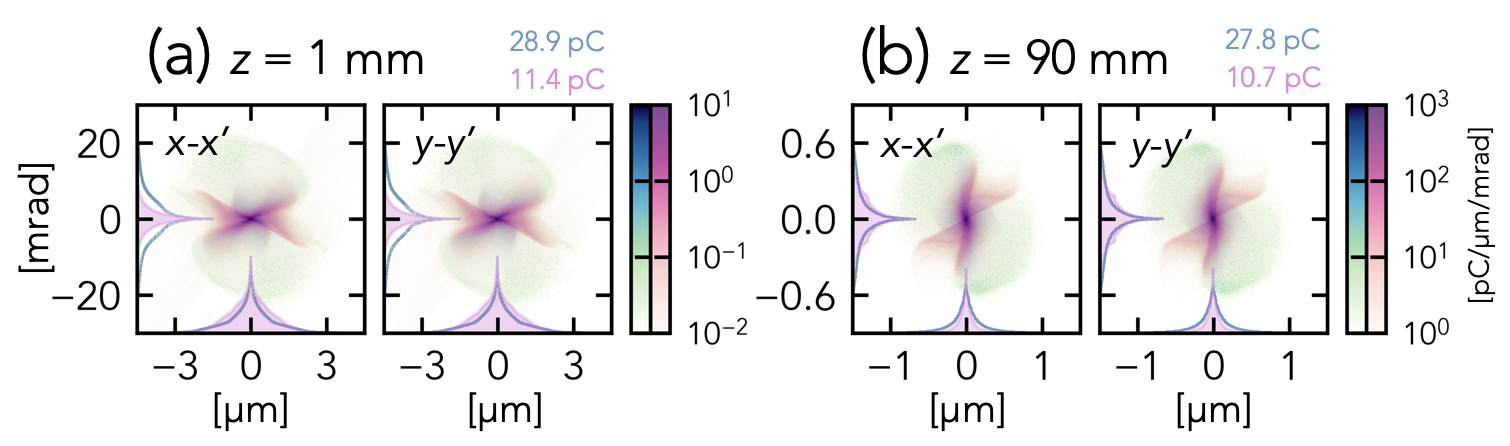}
	\caption{The transverse phase space of the trapped electrons at (a) $z = \SI{1}{mm}$ and (b) $z = \SI{90}{mm}$, the depletion length of the drive bunch. The trapped charge from each ionization source (He: blue, He$^+$: pink) is displayed above the sub-figures. Logarithmic colorbars are used to improve visibility of the strongly-peaked distributions.
 }
	\label{fig:RPLP_transverse_phasespace}
\end{figure}

Use of a RPLP necessitates a far-field intensity profile of a higher-order LG mode, as illustrated in the upper inset of Fig.~\ref{fig:overview}, in contrast to the fundamental Gaussian profile considered in previous studies. Unlike a fundamental Gaussian pulse whose ponderomotive force pushes ionized electrons away from the axis, the ``donut-like" intensity profile of the RPLP peaks off-axis generating a ponderomotive force which is radially-focusing for ionized electrons released within the peak intensity contour, and defocusing for those outside~\cite{KAWATA2005, GUPTA2007}. This acts to focus these electrons towards the axis which, in combination with the focusing wakefield, results in a strongly peaked spatial distribution of trapped electrons as seen in Fig.~\ref{fig:RPLP_transverse_phasespace}(a). This ponderomotive ``focusing" feature is not unique to radial polarization as it is due to the intensity distribution at focus, and hence would also be present for a linearly-polarized higher-order LG laser pulse. 

The radial nature of the electric field at focus leads to the generation of \textit{symmetric} transverse phase-space profiles of the injected electron bunch. This is demonstrated in Fig.~\ref{fig:RPLP_transverse_phasespace}(a), which shows the simulated transverse phase-space of the witness electron bunch after cessation of the injection process ($z = \SI{1}{mm}$). The distributions of trapped electrons in the $x-x'$ and $y-y'$ phase-spaces are identical, and remain so throughout propagation as demonstrated in Fig.~\ref{fig:RPLP_transverse_phasespace}(b) ($z = \SI{90}{mm}$). 

Around its focus, the tightly-focused RPLP generates a significant longitudinal field as demonstrated in the lower inset of Fig.~\ref{fig:overview} at $z = z_\mathrm{foc}$. This field acts to longitudinally accelerate the ionized electrons at their moment of release, but its magnitude is significantly smaller than that of the transverse laser field ($|E_x / E_z| \sim 50$) for the focusing geometry considered in this study. The initial momenta of the ionized electrons is therefore predominantly transverse, as in the linearly-polarized case, and no significant modification of the amount of charge trapped was observed for identical laser pulses with different polarization states but the same transverse intensity profile.

\begin{figure}[t]
	\centering
	\includegraphics[width = \linewidth]{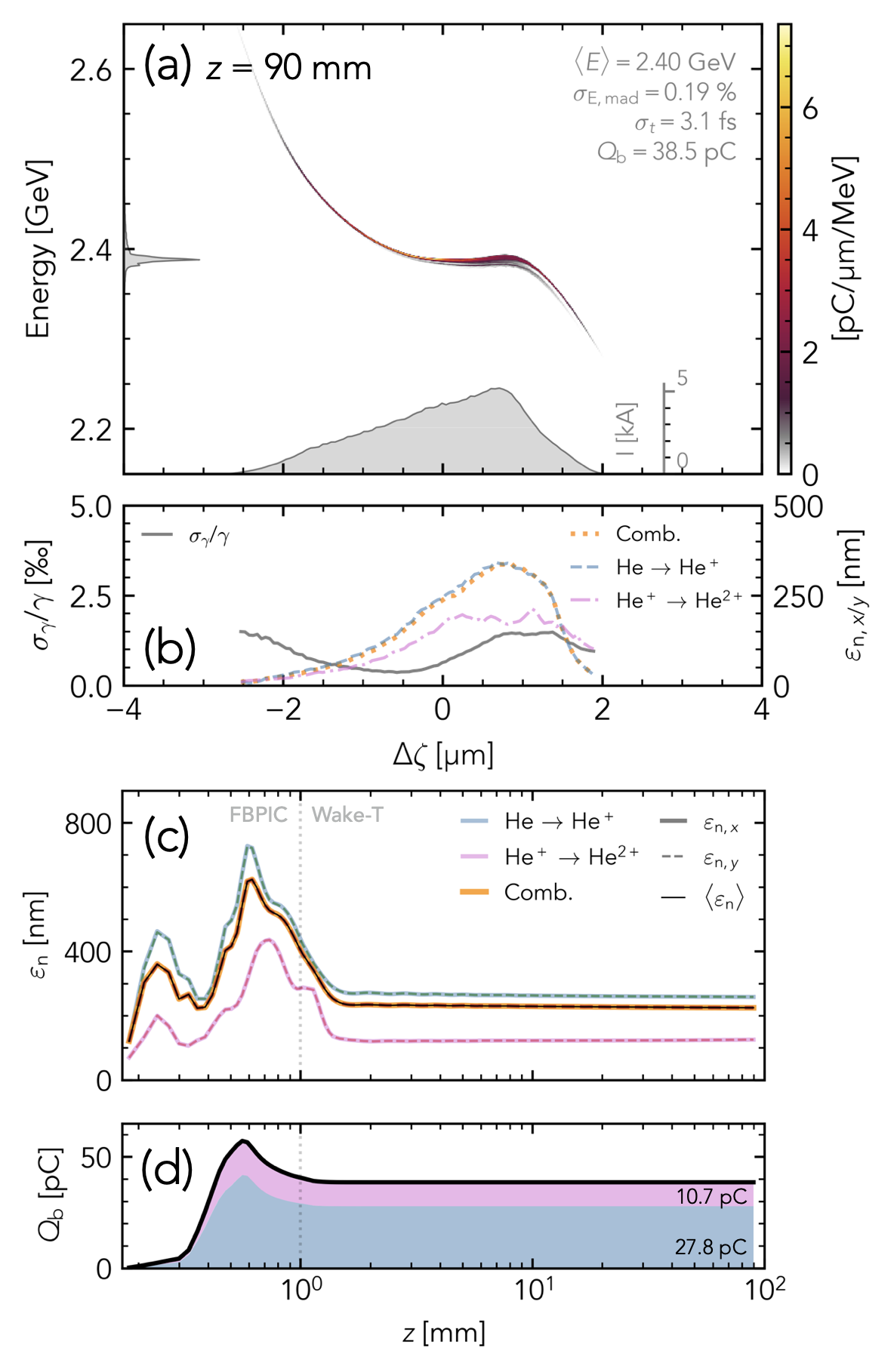}
	\caption{(a) Longitudinal phase space of the witness bunch after propagation over \SI{90}{mm}. (b) Slice energy spread (grey, solid) and normalised transverse slice emittance (orange, dotted). In (b), (c) and (d) the plotted quantities are separated by ionization source: He (blue); He$^+$ (pink); black lines represent calculations over the entire bunch. Evolution of the (c) projected normalised transverse emittance, and (d) charge, of the witness bunch over the entire propagation distance. In (c), thick solid (dashed) lines represent $\varepsilon_{\mathrm{n},x}$ ($\varepsilon_{\mathrm{n},y}$) for each ionization source; the thin black line represents $\langle \varepsilon_\mathrm{n} \rangle = \sqrt{\varepsilon_{\mathrm{n},x} \cdot \varepsilon_{\mathrm{n},y}}$ for the witness bunch. The vertical dotted line at $z = \SI{1}{mm}$ marks the handover between simulation codes.
 }
	\label{fig:RPLP_evo}
\end{figure}

Figure~\ref{fig:RPLP_evo} demonstrates the witness bunch parameters after depletion of the drive bunch at $z = \SI{90}{mm}$. The witness bunch has a charge of \SI{38.5}{pC} and an RMS duration of \SI{3.1}{fs}. The ultra-short bunch duration, despite the range of initial release positions of electrons in $\zeta$, is a key feature of the plasma photocathode scheme~\cite{Habib2023AnnPhys} and results in a high peak current of $I_\mathrm{pk} = \SI{5.2}{kA}$. The longitudinal phase space of the witness bunch, shown in Fig.~\ref{fig:RPLP_evo}(a), highlights the effectiveness of using an increased witness bunch charge to flatten the wakefield throughout acceleration. The witness bunch is accelerated to a mean electron energy of \SI{2.40}{GeV}, with a mean absolute deviation of \SI{4.5}{MeV} ($0.19\%$). The projected, normalised transverse emittance of the bunch is $\varepsilon_{\mathrm{n},x/y} = \SI{224}{nm}$ in both the $x-$ and $y-$planes. Figure~\ref{fig:RPLP_evo}(b) shows the longitudinal sliced parameters of the witness bunch, with a \textit{slice-averaged} relative energy spread of $0.11\%$ and transverse normalised emittance of $\varepsilon_{\mathrm{n,slice-av}} = \SI{212}{nm}$ in both planes.

\begin{figure}[b]
	\centering
	\includegraphics[width = \linewidth]{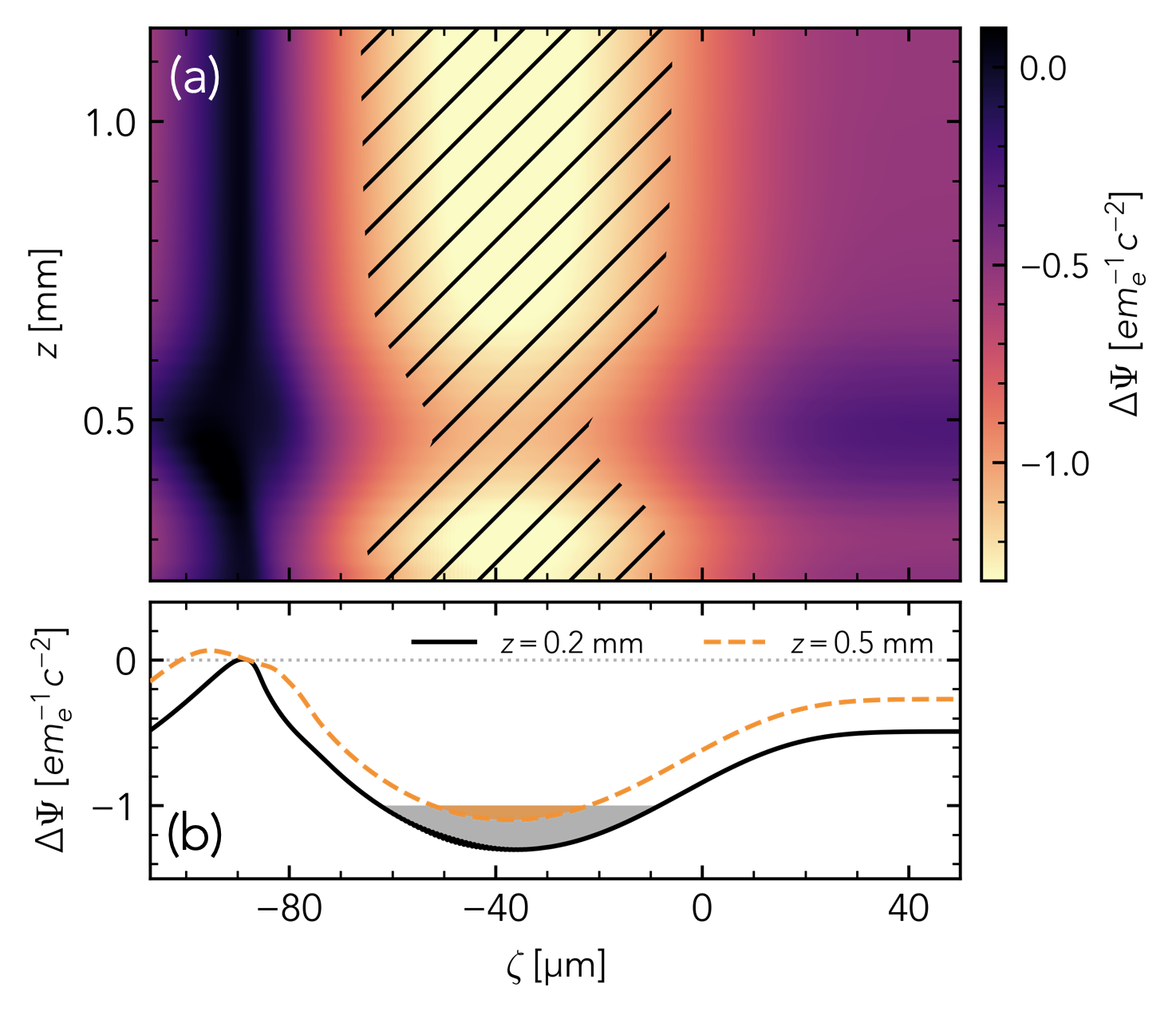}
	\caption{(a) Evolution of the on-axis trapping potential, $\Delta \Psi (\zeta, z) = \Psi_i - \Psi_f$, indicating the region for which released electrons can become trapped, $\Delta \Psi \lesssim -1$ (black, hatched). (b) Comparison of $\Delta \Psi$ just after trapping begins ($z = \SI{0.2}{mm}$, black) and as the wakefield becomes over-loaded ($z = \SI{0.5}{mm}$, orange dashed).
 }
	\label{fig:RPLP_psi}
\end{figure}

Figures~\ref{fig:RPLP_evo}(c) and (d) demonstrate the evolution of $\varepsilon_{\mathrm{n},x/y}$ and the trapped bunch charge, $Q_\mathrm{b}$, throughout propagation. Here, an electron is defined to be trapped when its longitudinal momentum exceeds $p_z \geq \SI{5}{MeV/}c$. Trapping first occurs at $z = \SI{170}{\micro m}$, before the RPLP focal position ($z_\mathrm{foc} = \SI{0.2}{mm}$), with the trapped bunch charge increasing continuously until $z \approx \SI{600}{\micro m}$. Figure~\ref{fig:RPLP_psi}(a) shows the evolution of the on-axis trapping region, $\Delta \Psi = \Psi_i - \Psi_f$, throughout the first \SI{1}{mm} of propagation. Here, $\Psi(\zeta) = - e m_e^{-1} c^{-2} \int E_{z0}(\zeta) d\zeta$ represents the on-axis wakefield potential normalised by the electron rest mass; $\Psi_i = \Psi(\zeta \approx \SI{-45}{\micro m})$ represents the value of the on-axis wakefield potential at the location where electrons are initially released; and $\Psi_f = \Psi(\zeta = \SI{-87}{\micro m}$) corresponds to the rear of the cavity at $z = \SI{0.2}{mm}$ and approximate longitudinal location of electrons that become trapped. In this way, $|\Delta \Psi|$ represents the maximum normalised energy an electron can gain as it slips backwards through the wakefield cavity --- its value for each electron will be reduced according to the trajectory each takes as $\Psi$ is maximised on-axis. The trapping region is therefore represented by $\Delta \Psi \lesssim -1$~\cite{Pak2010}, and corresponds to the region for which an electron born at rest within the wakefield gains enough longitudinal momentum to be trapped within the accelerating cavity. The boundaries of the on-axis trapping region are modified as electrons become trapped at the rear of the wakefield cavity, interfering with returning plasma electrons and briefly over-loading the wakefield. This causes the shape of the wakefield potential to be modified between $z \approx \SIrange{0.45}{0.6}{mm}$ [see Fig.~\ref{fig:RPLP_psi}(b)], reducing $\Delta \Psi$ for electrons released just prior to, and in, this region. This inhibits further trapping of electrons within the wakefield, as evidenced in Fig.~\ref{fig:RPLP_evo}(d). Beyond $z \approx \SI{0.6}{mm}$ the RPLP intensity has dropped sufficiently that it no longer ionizes the dopant helium. Similarly, $\varepsilon_{\mathrm{n},x/y}$ grows until $z = \SI{0.6}{mm}$, but subsequently rapidly decreases as initially trapped charge is lost as electrons at larger radii interact with the surrounding wakefield sheath and are accelerated away from the axis.  After $z \approx \SI{1.3}{mm}$ both $\varepsilon_{\mathrm{n},x/y}$ and $Q_\mathrm{b}$ stabilise and remain constant thereafter, indicating matched propagation of the bunch.

Also shown in Figs.~\ref{fig:RPLP_evo}(b) and (c) are the sliced and projected (respectively) normalised transverse emittances of electrons originating from the two ionization levels of helium. The total bunch charge of $Q_b = \SI{38.5}{pC}$ is split, with \SI{27.8}{pC} originating from the ionization of neutral He and \SI{10.7}{pC} from ionization of He$^+$. Interestingly, the contribution from He$^+$, shown in pink, has a consistently lower $\varepsilon_{\mathrm{n},x/y}$ than the contribution from neutral He (blue). 

\begin{figure}[t]
	\centering
	\includegraphics[width = \linewidth]{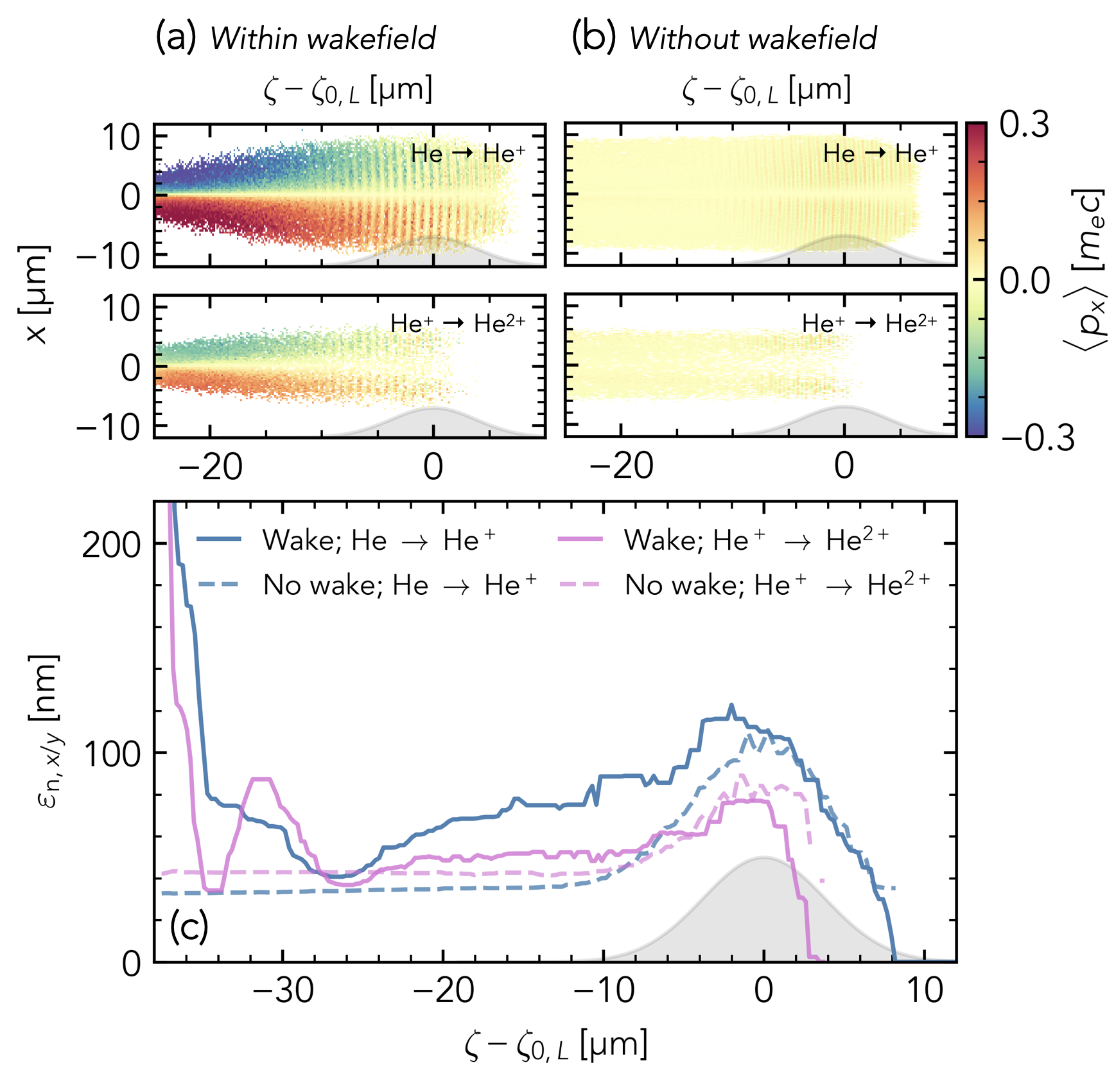}
	\caption{Evolution of the average transverse momentum, $\langle p_x \rangle$, of electrons ionized by the RPLP both (a) with and (b) without the presence of the wakefield. (c) Evolution of the transverse normalised slice emittance of electrons released from ionization of He (blue) and He$^+$ (pink) within a single simulation timestep, both with (solid) and without (dashed) the wakefield. Electrons slip backwards (towards $-\zeta$) before being captured at the rear of the wakefield cavity ($\zeta - \zeta_{0,L} \approx \SI{-38}{\micro m}$). The longitudinal profile of the RPLP is represented by the grey shaded region. In (c), a median filter has been applied to the data to remove high frequency oscillations due to the laser field, whose effect can clearly be seen in (a) and (b). This timestep corresponds to $z = \SI{0.2}{mm}$. 
 }
	\label{fig:RPLP_wake_emit_evo}
\end{figure}

Electrons released from the ionization of He$^+$ are born in regions of higher transverse --- and also proportionally higher longitudinal --- laser field than those released from neutral helium due to their higher ionization threshold. This increases the transverse momentum imparted to the electron by the laser field, resulting in a larger temperature --- or thermal emittance, $\varepsilon_{\mathrm{n,th}}$ --- at ionization. However, within the wakefield the emittance of electrons released from ionization of He$^{+}$ is consistently smaller than for those from He. This is likely caused by the fact that He$^+$ is ionized in a region of smaller spatial extent. Figures~\ref{fig:RPLP_wake_emit_evo}(a) and (b) show the average normalised transverse momentum of ionized electrons, $\langle p_x(\zeta, x) \rangle$, both within (a) and without (b) the wakefield, in a single simulation timestep. In (a), initially the transverse momenta are dominated by the transverse laser field, but after passage of the RPLP the transverse wakefield becomes the dominant effect, increasing $\langle p_x \rangle$ by an order of magnitude with increasing delay behind the laser pulse, with momenta directed towards the axis ($p_x < 0$ for $x > 0$ and vice versa). Figure~\ref{fig:RPLP_wake_emit_evo}(c) shows the evolution of the transverse emittance: ahead of the peak of the laser pulse, $\zeta - \zeta_{0,L} > 0$, the growth of the emittance with decreasing $\zeta$ is similar with and without the wakefield. However, behind the peak the transverse wakefield begins to dominate, focusing electrons within the wakefield [Fig.~\ref{fig:RPLP_wake_emit_evo}(a)] and causing slight emittance growth as the electrons become matched to the transverse fields. Additional focusing of the electrons towards the axis by the transverse wakefield helps to combat their expansion from the build-up of the space-charge force, minimising further emittance growth. When the electrons reach the rear of the wakefield cavity, $\zeta - \zeta_{0,L} \approx \SI{-38}{\micro m}$, the slice emittance grows significantly as electrons located off-axis interact with the returning plasma electron sheath.

\section*{Comparison to the linearly-polarized plasma photocathode}

\begin{figure}[t]
	\centering
	\includegraphics[width = \linewidth]{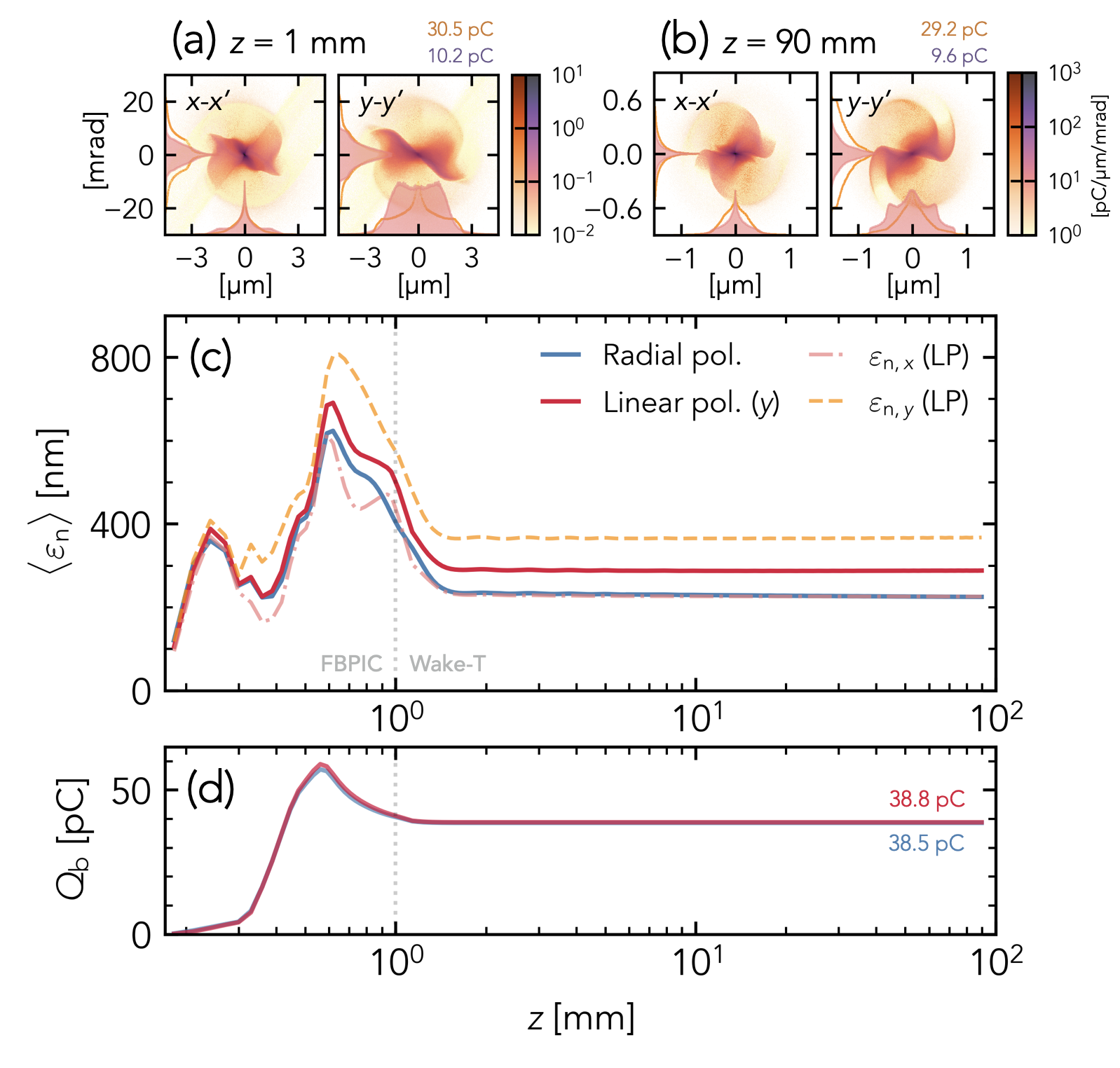}
	\caption{(a) and (b) The transverse phase space of the trapped charge at (a) $z = \SI{1}{mm}$ and (b) $z = \SI{90}{mm}$ for the linearly-polarized (in $y$) case with laser parameters matching those of the RPLP in Figs.~\ref{fig:overview}, ~\ref{fig:RPLP_transverse_phasespace} and~\ref{fig:RPLP_evo}. The orange (purple) colorbars represent electrons released via ionization of He (He$^+$), and the contribution to the total charge from each ionization level at the corresponding $z$-position is written above the plots. Comparison of the evolution of the (c) projected normalised transverse emittance, $\langle \varepsilon_{\mathrm{n}} \rangle = \sqrt{\varepsilon_{\mathrm{n},x} \cdot \varepsilon_{\mathrm{n},y}}$, and (d) bunch charge, for the radially- (blue) and linearly-polarized (red) cases. The emittance components in $x$ (pink, dot-dashed) and $y$ (orange, dashed) are also shown for the linearly-polarized case in (c).
 }
	\label{fig:LP_evo}
\end{figure}

In order to isolate the effect of the polarization state, the simulation was repeated for a linearly-polarized (in $y$) laser pulse with all other parameters unchanged, including the transverse intensity profile. Figure~\ref{fig:LP_evo}(a) shows the transverse phase-space of the trapped electrons following cessation of the injection process. In the $x-x'$ transverse phase-space --- the plane orthogonal to the polarization of the ionization pulse --- the distributions of the trapped electrons are strongly peaked around the axis, closely mirroring the results for the RPLP [Fig.~\ref{fig:RPLP_transverse_phasespace}(a)]. However, in the plane of polarization of the ionization pulse ($y-y'$) the spatial distribution is far wider and peaks off-axis. This broader distribution in $y-y'$ phase-space is maintained throughout acceleration [Fig.~\ref{fig:LP_evo}(b)], leading to a larger projected emittance $\varepsilon_{\mathrm{n},y} = \SI{367}{nm}$ for the final electron bunch. In the $x$-plane, the evolution of $\varepsilon_{\mathrm{n},x}$ closely mirrors that of (both planes of) the RPLP case, with $\varepsilon_{\mathrm{n},x} = \SI{225}{nm}$ at $z = \SI{90}{mm}$, as shown in Fig.~\ref{fig:LP_evo}(d). The larger emittance in the $y$-plane --- $\varepsilon_{\mathrm{n},y} / \varepsilon_{\mathrm{n},x} \approx 1.6$ --- is due to the contribution from the vector potential of the laser pulse at the moment of ionization, this time directed along $y$ alone due to the linear polarization of the field. The fact that $\varepsilon_{\mathrm{n},x}$ in the linearly-polarized case matches \textit{both} planes of the RPLP case indicates that for this high-charge, optimally-loaded regime of the plasma photocathode, the emittance growth due to mismatch of the initial ionized electron distribution to the transverse wakefield dominates the final emittance of the accelerated electron bunch, rather than their initial thermal emittance. The initial emittance asymmetry results in different mismatch in each plane of the bunch, inducing an asymmetric emittance growth during the matching process, resulting in a larger geometric average emittance, $\langle \varepsilon_{\mathrm{n}} \rangle = \sqrt{\varepsilon_{\mathrm{n},x} \cdot \varepsilon_{\mathrm{n},y}}$, than for the RPLP case. 

\begin{figure}[t]
	\centering
	\includegraphics[width = 0.9\linewidth]{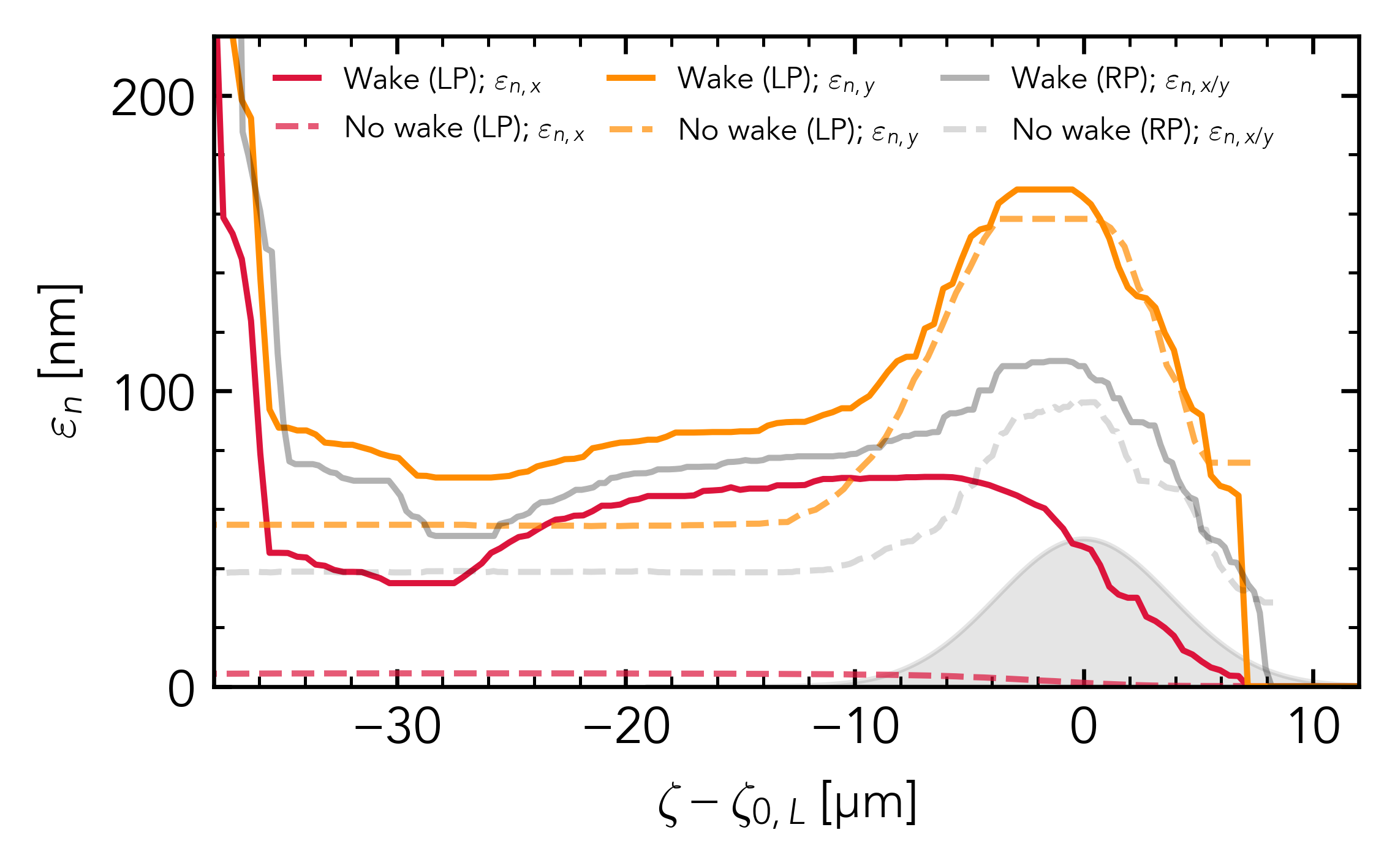}
	\caption{Evolution of the slice transverse normalised emittance of electrons released from ionization of He and He$^+$ by the linearly-polarized (LP) laser pulse within a single simulation timestep, both with (solid) and without (dashed) the wakefield. The emittance in the plane of polarization, $\varepsilon_{\mathrm{n},y}$, is shown in orange and represents electrons released via ionization of both He and He$^+$. The same data is shown in red for $\varepsilon_{\mathrm{n},x}$ (LP), and in grey for the RP pulse for comparison.
 }
	\label{fig:LP_wake_emit_evo}
\end{figure}

Simulations of the ionization of helium by both the radially- and linearly-polarized pulse \textit{without} the presence of the wakefield further demonstrated the difference in thermal emittance for the two polarizations. The radially-polarized pulse produced electrons from ionization of neutral He with a thermal emittance $\varepsilon_{\mathrm{n,th}} \approx \SI{33}{nm}$ after passage of the laser pulse, and $\varepsilon_{\mathrm{n,th}} \approx \SI{45}{nm}$ for ionization of He$^+$, as can be seen in Fig.~\ref{fig:RPLP_wake_emit_evo}(c). In both cases, the thermal emittances were symmetrical in $x$ and $y$. In contrast, a strong asymmetry was observed for the linearly-polarized laser pulse with thermal emittances of $\varepsilon_{\mathrm{n,th},x} = \SI{5}{nm}$ ($\SI{7}{nm}$) and $\varepsilon_{\mathrm{n,th},y} = \SI{50}{nm}$ ($\SI{75}{nm}$) for electrons released via ionization of the first (second) level of helium. The corresponding evolution of the emittance within a single simulation timestep for the linearly-polarized laser pulse is shown in Fig.~\ref{fig:LP_wake_emit_evo}. Within the plane of polarization the emittance evolution is similar both with and without the presence of the wakefield up until the peak of the laser pulse, $\zeta - \zeta_{0,L} = 0$. After this point, $\zeta - \zeta_{0,L} < 0$, the mismatch with the transverse wakefield leads to slight emittance growth in $\varepsilon_{\mathrm{n},y}$ relative to when there is no wakefield present. The maximum $\varepsilon_{\mathrm{n},y}$, occurring near the location of the peak intensity of the laser pulse as the electrons undergo quiver motion in the laser field, is larger than the equivalent for the RPLP [Fig.~\ref{fig:RPLP_wake_emit_evo}(c)] due to the increased transverse field amplitude of the linearly-polarized pulse --- for equal intensity, $E_\mathrm{max} (\mathrm{LP}) \simeq \sqrt{2} \cdot E_\mathrm{max} (\mathrm{RP})$. Out of the plane of polarization ($\varepsilon_{\mathrm{n},x}$) the difference with and without the presence of the wakefield is much more marked, highlighting the effect of the transverse mismatch and the severity of the emittance growth it induces. In the presence of the wakefield, the magnitude of the emittance asymmetry ($\varepsilon_{\mathrm{n},y}/ \varepsilon_{\mathrm{n},x}$) is much reduced after passage of the laser pulse, but the larger emittance in the plane of polarization remains.

The witness bunch charge is approximately equal for both polarizations, with $Q_b = \SI{38.8}{pC}$ for the linearly-polarized ionization pulse [Fig.~\ref{fig:LP_evo}(d)], confirming that the longitudinal field present at the focus of the RPLP has minimal impact on the electron trapping process. The energy spectrum is also very similar to the RPLP case, with the witness bunch accelerated to a mean electron energy of $\SI{2.38}{GeV}$, and a mean absolute deviation of $\SI{8.0}{MeV}$ ($0.33\%$).

\begin{figure*}[t]
	\centering
	\includegraphics[width = \linewidth]{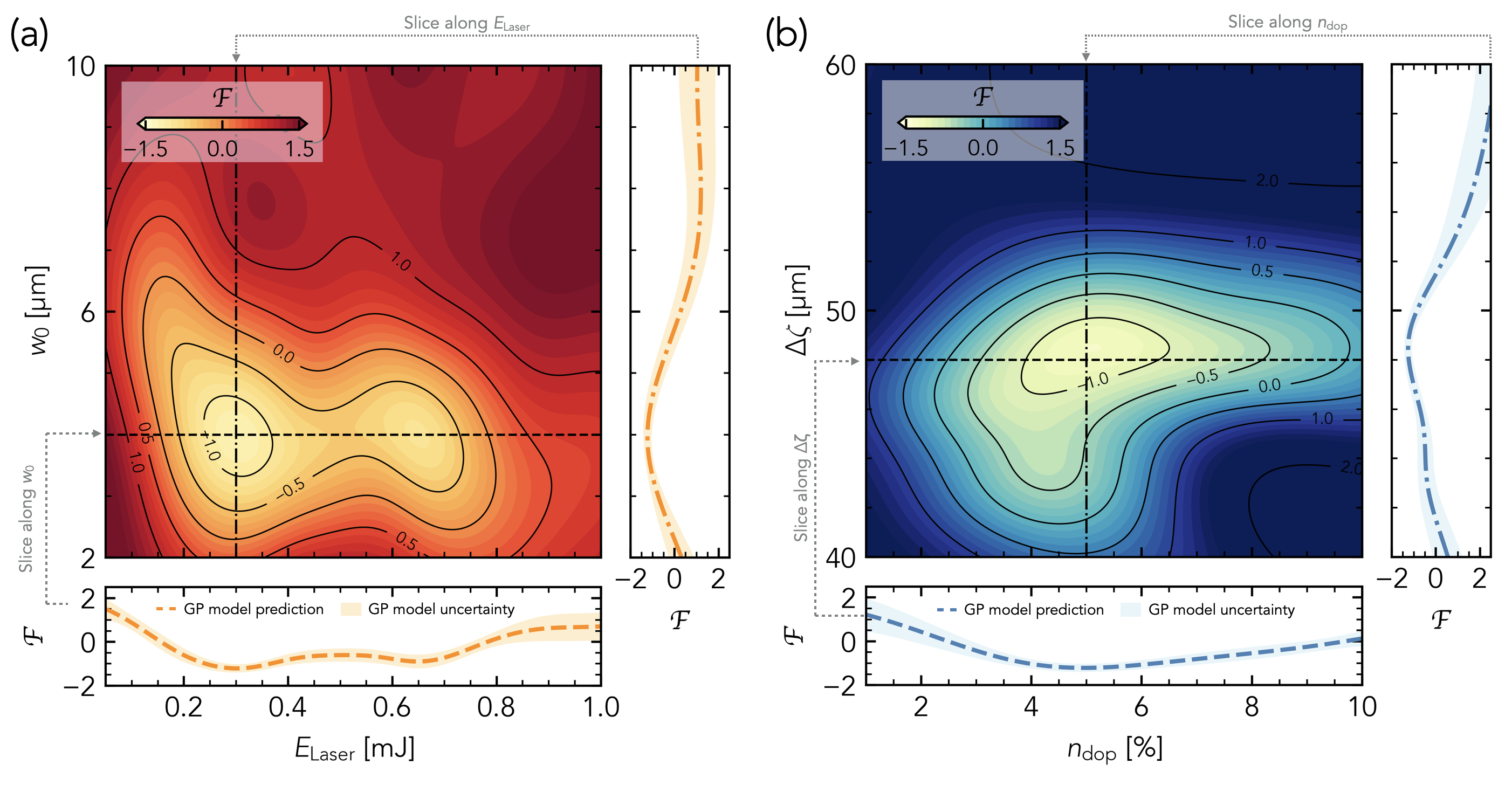}
	\caption{Visualisation of the GP model fitted to the results of the MOBO routine. Two slices along the 4-dimensional input parameter space are shown, taken at values close to those at which the minimum value of $\mathcal{F}$ was achieved: $E_\mathrm{Laser} = \SI{0.3}{mJ}$, $w_0 = \SI{4.0}{\micro m}$, $n_\mathrm{dop} = \SI{5}{\%}$, and $\Delta \zeta = \SI{48}{\micro m}$. Contour plots of the value of $\mathcal{F}$ predicted by the fitted GP model are shown as a function of (a) $w_0$ and $E_\mathrm{Laser}$, and (b) $\Delta \zeta$ and $n_\mathrm{dop}$. For each plot, GP model predictions and uncertainties are shown for the slices along the optimal input parameters. 
 }
	\label{fig:MOBO_slices}
\end{figure*}

\section*{Multi-objective Bayesian Optimisation}

In order to better understand the performance of the radially-polarized plasma photocathode scheme, a multi-objective Bayesian Optimisation (MOBO)~\cite{MOBOEmmerich2006,MOBODaulton2020} routine was used to explore the input parameter space. MOBO aims to find optimal trade-offs between multiple, often conflicting objectives to define the Pareto Front (PF) of the system --- the region of the parameter space for which one objective cannot be improved without deleterious effects on another. During performance of the MOBO routine, a surrogate Gaussian Process (GP) model~\cite{GPRasumussen2005} is continually updated to enable cheap, noise-free predictions of the outcome of simulations to be made and thus generate a search strategy to identify promising regions of the input parameter space to be explored. Here, the routine uses (noisy) expected hypervolume improvement~\cite{MOBODaulton2021} as the acquisition function. 

Four input parameters were varied: the total energy of the ionization laser pulse, $E_\mathrm{Laser}$; its spot size at focus, $w_0$; the fraction of helium dopant, $n_\mathrm{dop}$; and the relative delay between the centroid of the drive beam and ionization laser pulse, $\Delta \zeta$. The range over which these parameters were allowed to vary is summarised in Table~\ref{tab:table1}. The first three of these parameters were used to control the quantity of charge released into the wakefield via ionization of helium, while variation of the relative delay, $\Delta \zeta$, can be used to study the effects of timing fluctuations on the injection process.

The MOBO routine, performed using the \url{Optimas}~\cite{Optimas, LibEnsemble} Python library, aimed to simultaneously optimise three key metrics of the accelerated electron bunch: its charge, $Q_\mathrm{b}$; projected relative energy spread, $\sigma_{E,\mathrm{rel}} = \sigma_{E,\mathrm{mad}} / E_\mathrm{mean}$, where $\sigma_{E,\mathrm{mad}}$ is the RMS absolute deviation from the mean, $E_\mathrm{mean}$; and its geometric average transverse normalised emittance, $\langle \varepsilon_\mathrm{n} \rangle = \sqrt{\varepsilon_{\mathrm{n},x} \cdot \varepsilon_{\mathrm{n},y}}$. The value of an ``overall" optimisation metric, 

\begin{equation}
    \begin{split}
        \mathcal{F} = - \ln \Bigg[&\frac{E_{\mathrm{mean}}}{1\,\mathrm{GeV}} \cdot 
        \left(\frac{Q_b}{10\,\mathrm{pC}}\right)^{1/2} \cdot \\
        &\left(\frac{\sigma_{E,\mathrm{rel}}}{0.5\%}\right)^{-1} \cdot 
        \left(\frac{\langle \varepsilon_\mathrm{n}\rangle}{100 \,\mathrm{nm}}\right)^{-1} \Bigg],
    \end{split}
\end{equation}

\noindent which combines the three key output parameters in a way such that minimising $\mathcal{F}$ represents electron bunches with improved quality, was also tracked along with additional parameters such as the relative contributions of electrons released from the different ionization levels of helium to the final bunch charge. 

\begin{table}[t]
\centering
\begin{tblr}{
  width = \linewidth,
  colspec = {Q[300]Q[200]Q[200]Q[200]},
  cells = {c},
  hline{1,6} = {-}{0.08em},
  hline{2} = {-}{},
}
Parameter                         & Minimum & Maximum & Optimum \\
$E_\mathrm{Laser}\;[\mathrm{mJ}]$ & 0.05    & 1.0   & 0.3     \\
$w_0\;[\SI{}{\micro m}]$           & 2.0     & 10.0     & 4.0   \\
$n_\mathrm{dop}\;[\%]$            & 1.0     & 10.0  & 5.0    \\
$\Delta \zeta\;[\SI{}{\micro m}]$  & 40.0    & 60.0 & 48.0    
\end{tblr}
\caption{Summary of parameters used within the RPLP MOBO study and their minimum, maximum and optimum values.}
\label{tab:table1}
\end{table}

In total, 500 simulations for the RPLP setup were performed, with a further 200 simulations optimising the linearly-polarized case for comparison (see App.~\ref{app:modelling} for details of the start-to-end modelling workflow). Figure~\ref{fig:MOBO_slices} demonstrates the predictions of the resulting surrogate GP model after 500 simulations of the RPLP setup. Two contour plots are shown, representing the predicted value of $\mathcal{F}$ in 2D slices through the 4-dimensional input parameter space. The slices are taken at locations close to input parameters that generated the highest quality witness bunch, defined by minimising the value of $\mathcal{F}$: $E_\mathrm{Laser} = \SI{0.3}{mJ}$, $w_0 = \SI{4.0}{\micro m}$, $n_\mathrm{dop} = \SI{5}{\%}$, and $\Delta \zeta = \SI{48}{\micro m}$. While visualisation of the GP model in this way allows qualitative understanding of the stability of the input parameter space surrounding the optimal parameters, use of a single value, $\mathcal{F}$, to represent the quality of the outcome does not permit easy interpretation of correlations with individual witness bunch parameters. 

\subsection{Stability}

\begin{figure}[t]
	\centering
	\includegraphics[width = \linewidth]{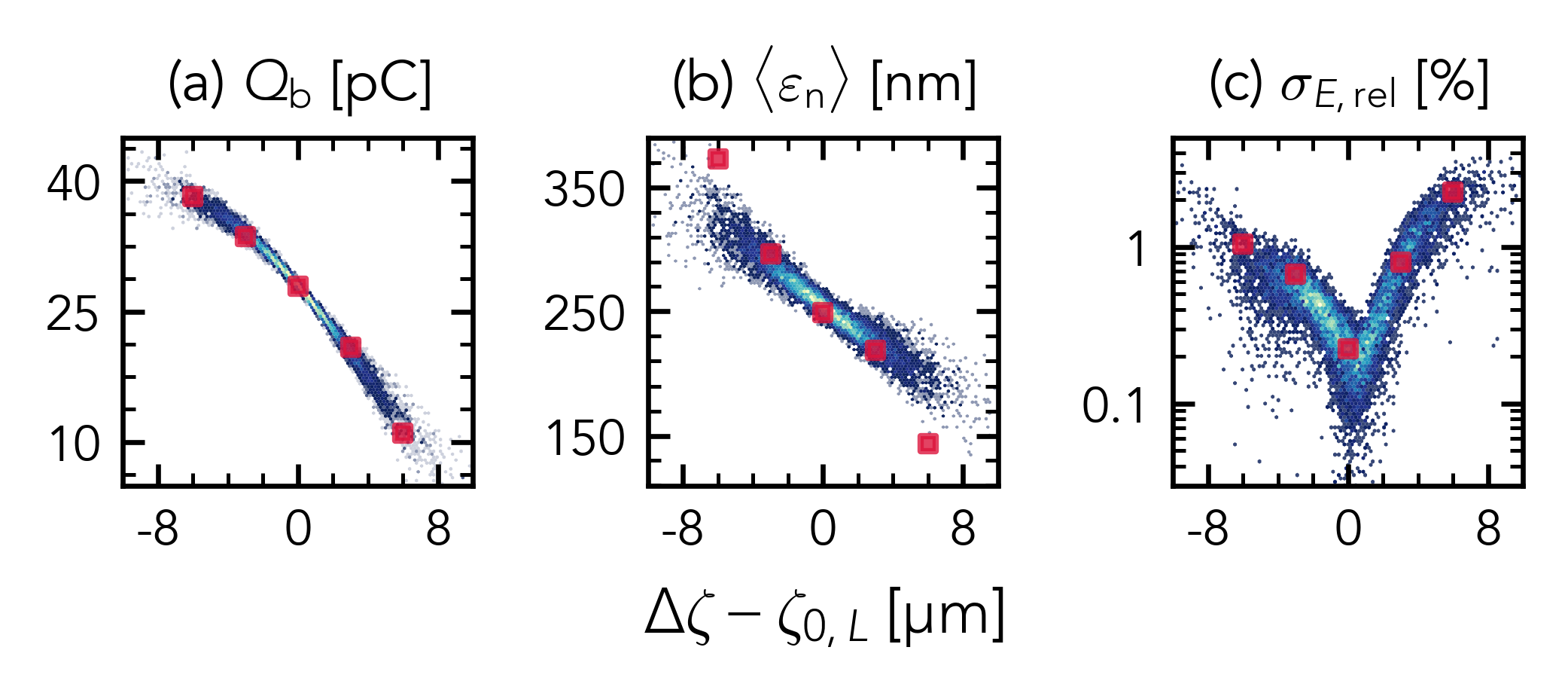}
	\caption{Distribution of witness bunch parameters resulting from MC sampling of the surrogate GP model when an RMS timing fluctuation of $\pm\SI{10}{fs}$ around the optimum is introduced. Red data points represent results from PIC simulations when varying $\Delta \zeta$ but keeping all other input parameters fixed. $\zeta_{0,L}$ is the location of the RPLP centroid.
 }
	\label{fig:GP_MC}
\end{figure}

To better quantify the stability of the input parameter space around the optimum, Monte-Carlo (MC) sampling ($N_\mathrm{samples} = 10^4$) of the surrogate model was performed. Typical experimental fluctuations were added to the optimal input parameters, sampled according to a normal distribution centred on the optimal value with a chosen relative width. The relative magnitude of the fluctuations were: $E_\mathrm{Laser} = 3\%$, $w_0 = 10\%$, $n_\mathrm{dop} = 10\%$, and $\Delta \zeta = 4\%$, chosen to represent typical fluctuations observed in experiments. Initially, each input parameter was varied individually, while keeping the others fixed to isolate the contributions to the fluctuations of the witness parameters. The dominant source of fluctuations was found to be the relative delay between the drive bunch and ionization laser pulse, $\Delta \zeta$, with a magnitude corresponding to a timing jitter of $\pm \SI{10}{fs}$, similar to state-of-the-art synchronisation systems~\cite{Schulz2015Timing, shalloo2015synchronisation, Christie2024}. The resulting witness bunch parameters varied over the range $Q_\mathrm{b} = \SI{27.3(6.7)}{pC}$, $\langle \varepsilon_\mathrm{n} \rangle = \SI{256(33)}{nm}$ and $\sigma_{E,\mathrm{rel}} = \SI{0.56(0.48)}{\%}$, where the quoted fluctuations represent the standard deviation. The variations of the key witness bunch parameters resulting from the MC sampling is shown in Fig.~\ref{fig:GP_MC}. Also shown are the results of PIC simulations (red data points) performed using the optimal input parameters but varying only $\Delta \zeta$ over the same range as tested with the MC sampling. The two agree well, confirming the quality of the surrogate model and its ability to make accurate predictions.

The witness bunch charge can be seen to be strongly correlated with the relative delay of the ionization laser pulse [Fig.~\ref{fig:GP_MC}(a)], with an increased bunch charge at smaller delays. This is because the ionization pulse centroid moves closer to the wakefield potential minimum --- $\zeta \approx \SI{-38}{\micro m}$ [Fig.~\ref{fig:RPLP_psi}] --- maximising $|\Delta \Psi|$ for the released electrons. However, this has two subsequent effects; the transverse emittance of the bunch increases and so does its relative energy spread as demonstrated in Figs.~\ref{fig:GP_MC}(b) and (c) respectively. The increase in bunch charge causes the wakefield to become locally \textit{over}-loaded, causing an energy chirp to be imparted to the witness bunch during acceleration. Conversely, for larger delays, electrons are released further back in the trapping region [Fig.~\ref{fig:RPLP_psi}(b)] and a smaller proportion of electrons gain enough longitudinal momentum to become trapped. This reduces the bunch charge and emittance, but now \textit{under}-loads the wakefield, again causing an increased energy spread. 

Such sensitivity to the relative timing between the drive bunch and laser pulse makes this scheme experimentally challenging, however there are ways in which this can be overcome. For example, operating the accelerator at a lower plasma density increases the size of the wakefield cavity and significantly relaxes the timing constraints. At the on-axis density used in this study, $n_0 = \SI{1e17}{cm^{-3}}$, a timing fluctuation of $\pm \SI{10}{fs}$ represents approximately $6\%$ of the duration of the wakefield cavity. Reducing the on-axis density to that proposed for future plasma-based accelerator facilities~\cite{HALHFFoster2023}, $n_0 = \SI{7e15}{cm^{-3}}$, increases the length of the trapping region such that the same fluctuation represents less than $2\%$ of the cavity, which would reduce the impact of such fluctuations. Furthermore, finding a combination of input parameters that enables optimal beam-loading where the position of the ionizing laser pulse corresponds to the minimum of the wakefield potential could further reduce the timing sensitivity thanks to the parabolic shape of the wakefield potential around its minimum~\cite{Habib2023AnnPhys}. We note also that recent work has demonstrated that active energy compression can reduce the shot-to-shot jitter of laser-accelerated electron bunches to the per mille level \cite{Winkler.2025}.

We note that the relative fluctuations in witness bunch parameters due to temporal jitter found in this study exceed those reported in Ref.~\cite{Habib2023AnnPhys} which were at most 5\% ($\sigma_{E,\mathrm{rel}}$), if not significantly less. However, the scenario modelled in Ref.~\cite{Habib2023AnnPhys} focused on minimising witness bunch emittance by using pulse intensities just above the ionization threshold of the dopant species, resulting in lower bunch charges ($\sim \SI{1}{pC}$). Furthermore, in that work the simulations considered witness acceleration over only \SI{8}{mm}, and not until drive bunch depletion. As discussed previously, the lower bunch charge precludes loading of the wakefield and causes the energy spread of the witness bunch to grow significantly during propagation, which will result in larger energy fluctuations by the end of the accelerator. To provide a point of comparison with this earlier work, the RPLP simulation results obtained in this work were sampled in a region of the parameter space that provided low-charge and ultra-low emittance witness bunches --- achieved by minimising the ionizing laser pulse energy and spot size. The MC procedure described above yielded, for this working point, witness bunch parameters of $Q_\mathrm{b} = \SI{2.8(1.4)}{pC}$, $\langle \varepsilon_\mathrm{n} \rangle = \SI{64(11)}{nm}$ and $\sigma_{E,\mathrm{rel}} = \SI{2.5(0.6)}{\%}$ when subject to the same temporal jitter as the optimally-loaded working point. The transverse emittance values are competitive with those found in previous studies~\cite{Hidding2012PRL, Hidding2014, Schroeder2014, Habib2023, Habib2023AnnPhys}, despite the increased thermal emittance in both transverse planes introduced by the field of the RPLP. However, the large relative energy spread and energy jitter obtained for the low-charge case would be challenging for further beam transport. The high-charge, optimally beam-loaded regime identified in this study would reduce problems of this kind.

We note also that the use of a higher-charge ``escort" bunch to locally flatten the longitudinal wakefield in the region of the low-charge witness~\cite{Habib2023} not only significantly increases experimental complexity but its generation will likely also be subject to similar fluctuations as those found in this study, inducing further fluctuations of the witness bunch parameters.

\subsection{Performance}

\begin{figure}[t]
	\centering
	\includegraphics[width = \linewidth]{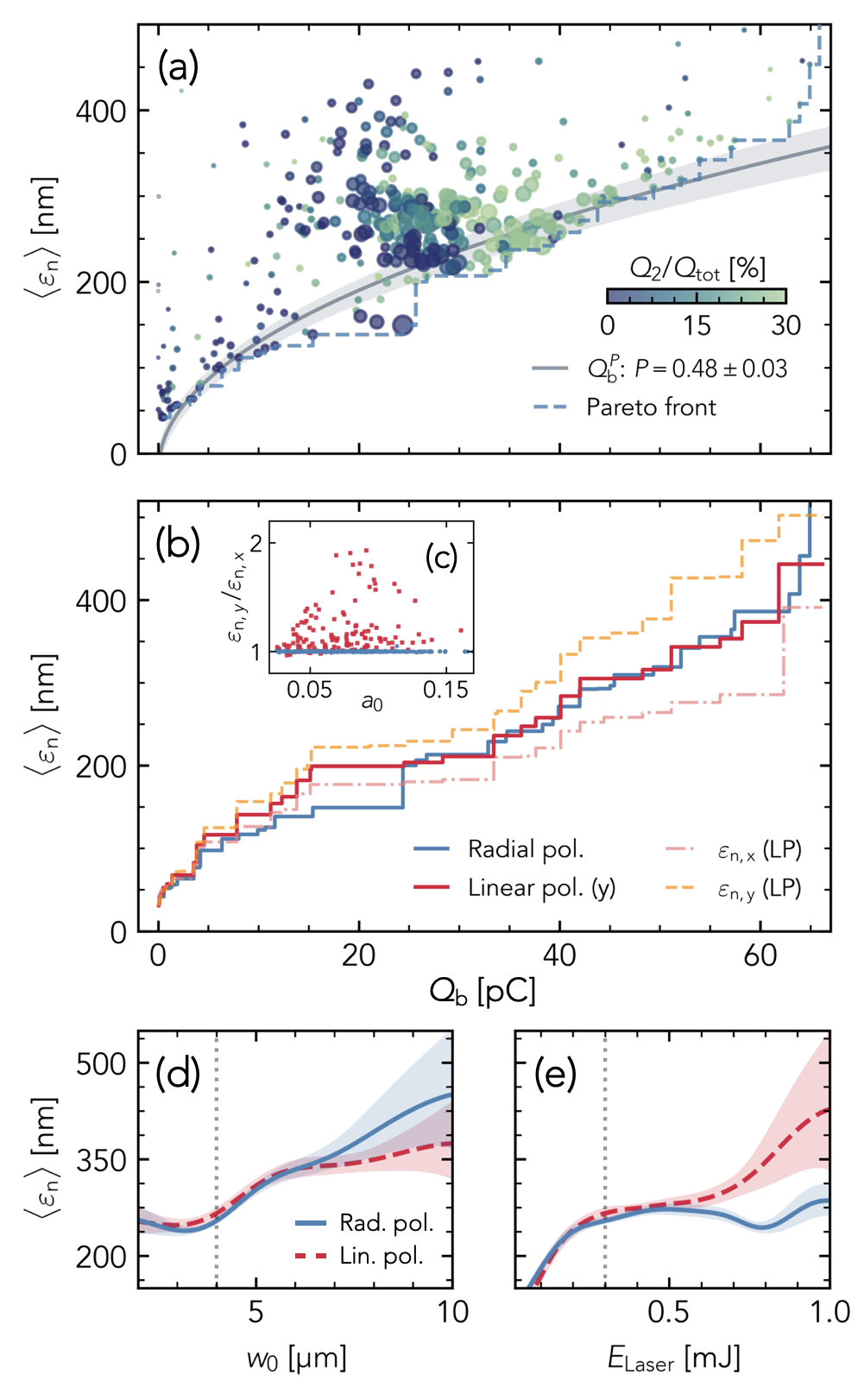}
	\caption{(a) Relation between $Q_b$ and $\langle \varepsilon_\mathrm{n} \rangle$ for the RPLP. The size of data points represents their value of $\mathcal{F}$, with larger points representing higher quality witness bunches. The colour of a data point indicates the fraction of the final witness bunch charge from ionization of He$^+$. The extracted PF is demonstrated by the dashed, blue line. (b) Comparison between the PFs for radial (blue) and linear polarization (red). The components of $\varepsilon_\mathrm{n}$ in the $x-$ (pink, dot-dashed) and $y-$planes (orange, dashed) are also shown for linear polarization. (c) $\varepsilon_{\mathrm{n},y} / \varepsilon_{\mathrm{n},x}$ as a function of ionization pulse normalised vector potential, $a_0$. (d) and (e) Mean (line) and standard error (shaded region) of predictions from the surrogate GP model demonstrating the variation in $\langle \varepsilon_\mathrm{n} \rangle$ with (d) $w_0$ and (e) $E_\mathrm{Laser}$ while keeping all other parameters fixed at their optima. The dashed grey vertical lines represent the optimal values of the laser parameters, as defined in Table~\ref{tab:table1}.
 }
	\label{fig:PFs}
\end{figure}

Further understanding of the ultimate performance of the radially-polarized plasma photocathode can be gleaned by exploring the relationship between bunch charge and emittance. Maximising bunch charge, and consequently peak current, while minimising transverse emittance is crucial for maximising the 6D brightness of the electron bunch. Generation of a high-brightness electron bunch using a traditional photocathode is in-part governed by a fundamental limitation relating the charge of the bunch to its emittance as a result of space-charge forces, an effect that is particularly dominant before electrons gain significant longitudinal momentum. A similar relation can be extracted for the plasma photocathode from the results of the MOBO routine. An example of this is presented in Fig.~\ref{fig:PFs}(a), which shows the relation between the final projected normalised average transverse emittance of the accelerated witness bunch, $\langle \varepsilon_\mathrm{n} \rangle$, and its charge, $Q_b$, for a radially-polarized ionization pulse. The Pareto front, indicated by the dashed blue line, represents the minimum achievable emittance of the witness bunch as a function of the bunch charge. A power law can be fitted to extract the relationship between the two: a fit performed using Bayesian linear regression indicated a power of \SI{0.48(0.03)}{}, where the uncertainty represents the $95\%$ confidence interval, therefore consistent with a scaling of $\langle \varepsilon_\mathrm{n} \rangle \propto \sqrt{Q_\mathrm{b}}$ in this regime. While a separate study would be required for alternative drive bunch parameters, this scaling indicates $\SI{}{\micro m}$-scale transverse emittances could be achieved with the radially-polarized plasma photocathode scheme when generating nC-class bunches, matching those required by future plasma-based linear collider designs~\cite{HALHFFoster2023}.

Figure~\ref{fig:PFs}(b) compares the extracted PFs for the radially- and linearly-polarized cases. The PF is very similar for both, indicating that in this regime the final emittance of the witness bunch is dominated by its evolution during the trapping and matching phase --- defined by the properties of the wakefield and therefore consistent in both cases --- and not by its thermal emittance at the point of ionization. Also shown are the corresponding PFs for the $x-$ and $y-$components of the emittance in the linearly-polarized case, demonstrating its asymmetry. Figure~\ref{fig:PFs}(c) shows the emittance asymmetry at $z = \SI{90}{mm}$ as a function of the ionization laser pulse normalised vector potential, $a_0 = (e \lambda E_\mathrm{max}) / (2 \pi m_e c^2)$, where $E_\mathrm{max}$ is the maximum of the laser pulse field. While the emittance remains symmetric over the entire range of $a_0$ simulated in the radially-polarized case (blue), the asymmetry grows with increasing $a_0$ for the linearly-polarized case (red), illustrating a key advantage of the radially-polarized scheme with the increased laser intensity necessary for high-charge operation.

It is important to note that while the PFs are very similar for both polarization states, the input parameters required to achieve minimal emittance can differ. When keeping the laser pulse parameters fixed and only changing its polarization, the average transverse emittance of the bunch is reduced with a RPLP when operating in the regime of optimal beam-loading of the wakefield, as shown previously in Fig.~\ref{fig:LP_evo}. Figures~\ref{fig:PFs}(d) and (e) demonstrate this effect over the entire range of ionizing laser pulse parameters considered. The figure shows the variation of the witness bunch emittance with (d) spot size and (e) pulse energy for the radially- and linearly-polarized ionization pulse while keeping all other parameters fixed at their optimal values. In the region of the parameter space around which the witness bunch optimally loads the wakefield and maximises its 6D brightness, indicated by the dashed grey lines, the average transverse emittance of the witness bunch is improved when using a radially-polarized ionizing laser pulse compared to linear polarization.

At the optimal input parameters found using the MOBO routine, the resulting witness bunch had a projected 6D brightness of $B_\mathrm{6D,n}\,(\mathrm{p}) = \SI{1.1e17}{}\mathrm{A m^{-2} \,0.1\%^{-1}}$, while the slice brightness peaked at $B_\mathrm{6D,n}\,(\mathrm{s}) = \SI{5.8e17}{}\mathrm{A m^{-2}\, 0.1\%^{-1}}$. For effective FEL generation, the dimensionless Pierce parameter $\rho$ is constrained by the relative energy spread of the bunch~\cite{DiMitri2015}, i.e. $\rho \geq \Delta E / E$. Here, the slice-averaged energy spread is found to be $\sigma_{E,\mathrm{rel, sl.}} = \SI{1.1e-3}{}$ with a minimum relative slice energy spread of $\approx \SI{4e-4}{}$, matching requirements of high-gain FELs. The minimum radiation wavelength produced by such a bunch can be calculated in the 1D limit via $\lambda_r \geq 4 \pi \varepsilon_\mathrm{n} / \gamma_0$~\cite{Huang2007}, where $\gamma_0$ is the relativistic Lorentz factor of the bunch. The ultra-low emittance and high-energies of the bunches produced in this scheme result in a minimum radiation wavelength of $\lambda_r \geq \SI{0.6}{nm}$. Lasing at shorter wavelenghts could be achieved by using a higher energy drive bunch to increase the energy of the witness bunch before drive depletion. A key advantage of operation in the optimally-loaded regime of the plasma photocathode is that increasing the acceleration length allows increase of the mean energy of the witness bunch without degradation of the energy spread or transverse emittance, thanks to the flattened longitudinal field gradient and matched propagation. The brightness values achieved in this study match, and even exceed, those currently produced at state-of-the-art X-FEL facilities~\cite{dimitri2014}, and demonstrate the ability of this scheme to transform the brightness of existing light sources with minimal impact on the footprint of the facility. For example, the drive bunch parameters chosen in this study closely match those provided by the FLASH facility at DESY~\cite{FLASHTiedtke2009} and could increase its 6D brightness by multiple orders of magnitude through the implementation of this scheme, while pushing the radiation wavelength below the nm-scale --- currently FLASH I/II operates at $B_\mathrm{6D,n}\,(\mathrm{p}) \sim \mathcal{O} (10^{15}\mathrm{A m^{-2}\, 0.1\%^{-1}})$, $B_\mathrm{6D,n}\,(\mathrm{s}) \sim \mathcal{O} (10^{16}\mathrm{A m^{-2}\, 0.1\%^{-1}})$, $\lambda_r = 4-6\,$nm. 

\section*{Conclusions}

Using start-to-end simulations we have demonstrated the high-charge regime of plasma photocathode operation, enabling optimal beam-loading from a single injection event to simultaneously minimise the projected energy spread and maximise the 6D brightness of the witness bunch. Use of a radially-polarized ionizing laser pulse enables generation of witness bunches with symmetric emittance in the $x-$ and $y-$planes, improving the 6D brightness of the witness bunch compared to an equivalent linearly-polarized ionizing laser pulse in the optimally-loaded regime. Furthermore, having identical emittance in both transverse planes eliminates the risk of emittance mixing during acceleration and eases beam transport and matching after the accelerator, while offering improved radiation generation opportunities, for example in helical undulators. 

In the high-charge regime of the plasma photocathode scheme, trapping of electrons released from higher ionization levels of the dopant species was shown to reduce the emittance of the bunch due to the reduced spatial extent of the ionized electrons, in contrast to behaviour outside of the wakefield environment. This is promising for controlled electron injection schemes relying on localised, selective ionization within quasi-linear laser-wakefield accelerator (LWFA) stages~\cite{Bourgeois2013, Yu2014}. In particular, it suggests the plasma photocathode scheme could be used to generate high-quality electron bunches within LWFA stages driven by a train of laser pulses~\cite{Hooker2014MPLWFA, Jakobsson2021} as the reduced peak vector potential ($a_0 \ll 1$) relative to single-pulse LWFA could still allow operation at low ionization laser pulse energies.

Multi-objective Bayesian Optimisation was used to explore the stability and ultimate performance of the high-charge regime, indicating particular sensitivity to the relative timing of the drive bunch and ionizing laser pulse. Extraction of the Pareto front relating the bunch charge to its average transverse normalised emittance indicated a scaling $\langle \varepsilon_\mathrm{n} \rangle \propto \sqrt{Q_\mathrm{b}}$, suggesting the generation of $\SI{}{\micro m}$-scale emittances for nC-scale bunch charges as required for future plasma-based facilities.

\begin{acknowledgments}
The authors would like to ackowledge useful discussions with R. D'Arcy. This work was supported by the UK Science and Technology Facilities Council (STFC UK), Grants No. ST/V001655/1 and agreement 9253 S2 2021 010; the Engineering and Physical Sciences Research Council, Grants No. EP/R513295/1 and No. EP/V006797/1. Computing resources provided by STFC Scientific Computing Department’s SCARF cluster. This research was funded in whole, or in part, by EPSRC and STFC, which are Plan S funders.
\end{acknowledgments}


\appendix

\section{Start-to-end modelling}
\label{app:modelling}

Modelling of the first \SI{1}{mm} of the interaction was performed with the \url{FBPIC} (v.\url{0.26.0}) particle-in-cell code using $N_\mathrm{m} = 3$ azimuthal modes. The injection simulations corresponding to results in Figs.~\ref{fig:overview}--\ref{fig:LP_wake_emit_evo} were performed in the lab frame, with 48 particles per cell (PPC) for each species: pre-ionized plasma electrons and the dopant neutral helium. A non-evolving background of ions was assumed as there was no discernable difference in wakefield or witness bunch parameters seen when including mobile hydrogen ions. The simulation domain measured $\SI{60}{\micro m} \times \SI{157}{\micro m}$ ($r \times z$), with $\Delta r = \SI{0.5}{\micro m}$ and $\Delta z = \SI{0.02}{\micro m}$ (i.e. $\lambda / 40$). The injected witness bunch was typically formed of $\sim \SI{4e6}{}$ macro-particles (depending on its charge), while the Gaussian drive bunch was modelled using \SI{1e6} equal weight macro-particles.

At the end of the ionization simulation, the drive and witness bunch distributions were written to file and then loaded into the extended propagation simulation in \url{Wake-T} (v.\url{0.8.0}). The witness bunch distributions were separated according to the ionization source, either neutral He or He$^+$, to enable tracking of the separate components. The \url{Wake-T} simulation used an identical size of simulation domain with $\Delta r = \SI{0.25}{\micro m}$ and $\Delta z = \SI{0.14}{\micro m}$. 

The propagation distance at which the handover between the ionization (\url{FBPIC}) and extended propagation (\url{Wake-T}) simulations was performed was scanned between $z_\mathrm{h} = 0.6 - 2.0$\,mm. With $z_\mathrm{h} = \SI{0.6}{mm}$, the final witness bunch charge after drive bunch depletion at $z = \SI{90}{mm}$ was lower due to electrons released from ionization of helium later in the simulation not yet meeting the ``trapped" criterion ($p_z \geq \SI{5}{MeV}/c$) by the handover position, $z_\mathrm{h}$. For handover positions $z_\mathrm{h} \geq \SI{1.0}{mm}$ the witness bunch charge (and all other properties) was consistent at $z = \SI{90}{mm}$ in all cases, and therefore a handover position of $z_\mathrm{h} = \SI{1.0}{mm}$ was selected to minimise the duration of the ionization simulation.

The extended propagation simulation performed using \url{Wake-T} was first benchmarked against a full 3D \url{HiPACE++}~\cite{Diederichs2022HiPACE++} (v.\url{24.03}) simulation using the same drive and witness bunches output by a test \url{FBPIC} simulation after \SI{1}{mm} of propagation. The benchmark \url{HiPACE++} simulation had a global grid resolution of $\Delta x = \Delta y = \SI{0.26}{\micro m}$, $\Delta z = \SI{0.08}{\micro m}$ with 4 PPC for plasma electrons and 1 PPC for (mobile) ions. Around the witness bunch, a region of radius \SI{12.5}{\micro m} was further resolved using mesh refinement, with an increased resolution of $\Delta x = \Delta y = \SI{0.032}{\micro m}$ with 256 PPC for plasma electrons and 64 PPC for ions. For both the \url{Wake-T} and \url{HiPACE++} simulations, an adaptive timestep ($\Delta t$) was used with value corresponding to $\omega_\beta \Delta t = 2 \pi / 20$, where $\omega_\beta = \omega_{p0} / \sqrt{2 \gamma}$ with $\omega_{p0}$ the plasma angular frequency and $\gamma$ an average of Lorentz factors of the slowest particles in both the drive and witness beams. The benchmark \url{HiPACE++} simulation took approximately 200 minutes to perform using $4\times$ NVIDIA A100 (40GB) GPUs, while the \url{Wake-T} simulation took approximately 20 minutes on a single AMD EPYC 7502 CPU core. A comparison of the witness bunch parameters for the two simulations after \SI{90}{mm} propagation are shown in Table~\ref{tab:waket_vs_hipace} and show excellent agreement, justifying the use of \url{Wake-T} for extended propagation simulations thanks to the reduced computational requirements.

\begin{table}[!b]
\centering
\begin{tblr}{
  width = \linewidth,
  colspec = {Q[400]Q[300]Q[300]},
  cells = {c},
  hline{1,8} = {-}{0.08em},
  hline{2} = {-}{},
}
Parameter                                   & HiPACE++ & Wake-T \\
$E_\mathrm{mean}\;[\mathrm{GeV}]$           & 2.56     & 2.57   \\
$\sigma_{E,\mathrm{rel}}\;[\%]$             & 3.8      & 3.9    \\
$I_\mathrm{pk}\;[\mathrm{kA}]$              & 3.1      & 3.1    \\
$Q_\mathrm{b}\;[\mathrm{pC}]$               & 17.7     & 17.6   \\
$\varepsilon_{\mathrm{n},x}\;[\mathrm{nm}]$ & 219      & 221    \\
$\varepsilon_{\mathrm{n},y}\;[\mathrm{nm}]$ & 220      & 222    
\end{tblr}
\caption{Comparison between witness bunch parameters from a test simulation after $\SI{90}{mm}$ propagation using HiPACE++ (including mesh refinement and ion motion) and Wake-T.}
\label{tab:waket_vs_hipace}
\end{table}

Ionization simulations performed with \url{FBPIC} during the MOBO routine (Figs.~\ref{fig:MOBO_slices}--\ref{fig:PFs}) used the boosted frame ($\gamma_\mathrm{boost} = 2)$ to reduce computational time. Convergence tests demonstrated consistent witness bunch parameters between the lab and boosted frames, although with fewer macroparticles representing the witness bunch in the boosted frame --- on average, the bunch was represented by $\sim \SI{5e5}{}$ macroparticles. The size of the simulation domain, resolution used and PPC matched those of the lab frame simulation, although these simulations also included mobile plasma ions. 

Handover between the various steps of the simulation workflow were handled by the \url{ChainEvaluator} function available within the \url{Optimas} (v.\url{0.6.0}) Python package. For each trial (i.e. set of input variables), a single NVIDIA A100 (40 GB) GPU was used to perform the ionization simulation in \url{FBPIC}, followed by the extended propagation simulation in \url{Wake-T} performed on a single AMD EPYC 7502 CPU core. A single start-to-end simulation took approximately 100 minutes to complete, with 80 minutes required for injection modelling and 20 minutes for extended propagation. Four separate simulations could be run in parallel, with the entire optimisation routine performed in approximately 200 hours.

The version of the \url{LASY} Python package used in this study was v.\url{0.4.0}.

\bibliography{references}

\begin{thebibliography}{49}%
\makeatletter
\providecommand \@ifxundefined [1]{%
 \@ifx{#1\undefined}
}%
\providecommand \@ifnum [1]{%
 \ifnum #1\expandafter \@firstoftwo
 \else \expandafter \@secondoftwo
 \fi
}%
\providecommand \@ifx [1]{%
 \ifx #1\expandafter \@firstoftwo
 \else \expandafter \@secondoftwo
 \fi
}%
\providecommand \natexlab [1]{#1}%
\providecommand \enquote  [1]{``#1''}%
\providecommand \bibnamefont  [1]{#1}%
\providecommand \bibfnamefont [1]{#1}%
\providecommand \citenamefont [1]{#1}%
\providecommand \href@noop [0]{\@secondoftwo}%
\providecommand \href [0]{\begingroup \@sanitize@url \@href}%
\providecommand \@href[1]{\@@startlink{#1}\@@href}%
\providecommand \@@href[1]{\endgroup#1\@@endlink}%
\providecommand \@sanitize@url [0]{\catcode `\\12\catcode `\$12\catcode `\&12\catcode `\#12\catcode `\^12\catcode `\_12\catcode `\%12\relax}%
\providecommand \@@startlink[1]{}%
\providecommand \@@endlink[0]{}%
\providecommand \url  [0]{\begingroup\@sanitize@url \@url }%
\providecommand \@url [1]{\endgroup\@href {#1}{\urlprefix }}%
\providecommand \urlprefix  [0]{URL }%
\providecommand \Eprint [0]{\href }%
\providecommand \doibase [0]{https://doi.org/}%
\providecommand \selectlanguage [0]{\@gobble}%
\providecommand \bibinfo  [0]{\@secondoftwo}%
\providecommand \bibfield  [0]{\@secondoftwo}%
\providecommand \translation [1]{[#1]}%
\providecommand \BibitemOpen [0]{}%
\providecommand \bibitemStop [0]{}%
\providecommand \bibitemNoStop [0]{.\EOS\space}%
\providecommand \EOS [0]{\spacefactor3000\relax}%
\providecommand \BibitemShut  [1]{\csname bibitem#1\endcsname}%
\let\auto@bib@innerbib\@empty
\bibitem [{\citenamefont {Pellegrini}\ \emph {et~al.}(2016)\citenamefont {Pellegrini}, \citenamefont {Marinelli},\ and\ \citenamefont {Reiche}}]{Pellegrini2016}%
  \BibitemOpen
  \bibfield  {author} {\bibinfo {author} {\bibfnamefont {C.}~\bibnamefont {Pellegrini}}, \bibinfo {author} {\bibfnamefont {A.}~\bibnamefont {Marinelli}},\ and\ \bibinfo {author} {\bibfnamefont {S.}~\bibnamefont {Reiche}},\ }\bibfield  {title} {\bibinfo {title} {The physics of x-ray free-electron lasers},\ }\href {https://doi.org/10.1103/RevModPhys.88.015006} {\bibfield  {journal} {\bibinfo  {journal} {Rev. Mod. Phys.}\ }\textbf {\bibinfo {volume} {88}},\ \bibinfo {pages} {015006} (\bibinfo {year} {2016})}\BibitemShut {NoStop}%
\bibitem [{\citenamefont {Seddon}\ \emph {et~al.}(2017)\citenamefont {Seddon}, \citenamefont {Clarke}, \citenamefont {Dunning}, \citenamefont {Masciovecchio}, \citenamefont {Milne}, \citenamefont {Parmigiani}, \citenamefont {Rugg}, \citenamefont {Spence}, \citenamefont {Thompson}, \citenamefont {Ueda}, \citenamefont {Vinko}, \citenamefont {Wark},\ and\ \citenamefont {Wurth}}]{Seddon_2017}%
  \BibitemOpen
  \bibfield  {author} {\bibinfo {author} {\bibfnamefont {E.~A.}\ \bibnamefont {Seddon}}, \bibinfo {author} {\bibfnamefont {J.~A.}\ \bibnamefont {Clarke}}, \bibinfo {author} {\bibfnamefont {D.~J.}\ \bibnamefont {Dunning}}, \bibinfo {author} {\bibfnamefont {C.}~\bibnamefont {Masciovecchio}}, \bibinfo {author} {\bibfnamefont {C.~J.}\ \bibnamefont {Milne}}, \bibinfo {author} {\bibfnamefont {F.}~\bibnamefont {Parmigiani}}, \bibinfo {author} {\bibfnamefont {D.}~\bibnamefont {Rugg}}, \bibinfo {author} {\bibfnamefont {J.~C.~H.}\ \bibnamefont {Spence}}, \bibinfo {author} {\bibfnamefont {N.~R.}\ \bibnamefont {Thompson}}, \bibinfo {author} {\bibfnamefont {K.}~\bibnamefont {Ueda}}, \bibinfo {author} {\bibfnamefont {S.~M.}\ \bibnamefont {Vinko}}, \bibinfo {author} {\bibfnamefont {J.~S.}\ \bibnamefont {Wark}},\ and\ \bibinfo {author} {\bibfnamefont {W.}~\bibnamefont {Wurth}},\ }\bibfield  {title} {\bibinfo {title} {Short-wavelength free-electron laser sources and science: a review},\ }\href {https://doi.org/10.1088/1361-6633/aa7cca} {\bibfield  {journal} {\bibinfo  {journal} {Reports on Progress in Physics}\ }\textbf {\bibinfo {volume} {80}},\ \bibinfo {pages} {115901} (\bibinfo {year} {2017})}\BibitemShut {NoStop}%
\bibitem [{\citenamefont {Tetsuya}(2019)}]{Tetsuya2019}%
  \BibitemOpen
  \bibfield  {author} {\bibinfo {author} {\bibfnamefont {I.}~\bibnamefont {Tetsuya}},\ }\bibfield  {title} {\bibinfo {title} {{Accelerator-based X-ray sources: synchrotron radiation, X-ray free electron lasers and beyond}},\ }\href@noop {} {\bibfield  {journal} {\bibinfo  {journal} {Phil. Trans. R. Soc. A.}\ }\textbf {\bibinfo {volume} {377}},\ \bibinfo {pages} {20180231} (\bibinfo {year} {2019})}\BibitemShut {NoStop}%
\bibitem [{\citenamefont {Shin}(2021)}]{Shin2021}%
  \BibitemOpen
  \bibfield  {author} {\bibinfo {author} {\bibfnamefont {S.}~\bibnamefont {Shin}},\ }\bibfield  {title} {\bibinfo {title} {New era of synchrotron radiation: fourth-generation storage ring},\ }\href {https://doi.org/10.1007/s43673-021-00021-4} {\bibfield  {journal} {\bibinfo  {journal} {AAPPS Bulletin}\ }\textbf {\bibinfo {volume} {31}},\ \bibinfo {pages} {21} (\bibinfo {year} {2021})}\BibitemShut {NoStop}%
\bibitem [{\citenamefont {Habib}\ \emph {et~al.}(2023{\natexlab{a}})\citenamefont {Habib}, \citenamefont {Heinemann}, \citenamefont {Manahan}, \citenamefont {Ullmann}, \citenamefont {Scherkl}, \citenamefont {Knetsch}, \citenamefont {Sutherland}, \citenamefont {Beaton}, \citenamefont {Campbell}, \citenamefont {Rutherford}, \citenamefont {Boulton}, \citenamefont {Nutter}, \citenamefont {Hewitt}, \citenamefont {Dickson}, \citenamefont {Karger}, \citenamefont {Litos}, \citenamefont {O'Shea}, \citenamefont {Andonian}, \citenamefont {Bruhwiler}, \citenamefont {Pretzler}, \citenamefont {Wilson}, \citenamefont {Sheng}, \citenamefont {Stumpf}, \citenamefont {Reichwein}, \citenamefont {Pukhov}, \citenamefont {Cary}, \citenamefont {Hogan}, \citenamefont {Yakimenko}, \citenamefont {Rosenzweig},\ and\ \citenamefont {Hidding}}]{Habib2023AnnPhys}%
  \BibitemOpen
  \bibfield  {author} {\bibinfo {author} {\bibfnamefont {A.~F.}\ \bibnamefont {Habib}}, \bibinfo {author} {\bibfnamefont {T.}~\bibnamefont {Heinemann}}, \bibinfo {author} {\bibfnamefont {G.~G.}\ \bibnamefont {Manahan}}, \bibinfo {author} {\bibfnamefont {D.}~\bibnamefont {Ullmann}}, \bibinfo {author} {\bibfnamefont {P.}~\bibnamefont {Scherkl}}, \bibinfo {author} {\bibfnamefont {A.}~\bibnamefont {Knetsch}}, \bibinfo {author} {\bibfnamefont {A.}~\bibnamefont {Sutherland}}, \bibinfo {author} {\bibfnamefont {A.}~\bibnamefont {Beaton}}, \bibinfo {author} {\bibfnamefont {D.}~\bibnamefont {Campbell}}, \bibinfo {author} {\bibfnamefont {L.}~\bibnamefont {Rutherford}}, \bibinfo {author} {\bibfnamefont {L.}~\bibnamefont {Boulton}}, \bibinfo {author} {\bibfnamefont {A.}~\bibnamefont {Nutter}}, \bibinfo {author} {\bibfnamefont {A.}~\bibnamefont {Hewitt}}, \bibinfo {author} {\bibfnamefont {A.}~\bibnamefont {Dickson}}, \bibinfo {author} {\bibfnamefont {O.~S.}\ \bibnamefont {Karger}}, \bibinfo {author} {\bibfnamefont {M.~D.}\ \bibnamefont {Litos}}, \bibinfo {author} {\bibfnamefont {B.~D.}\ \bibnamefont {O'Shea}}, \bibinfo {author} {\bibfnamefont {G.}~\bibnamefont {Andonian}}, \bibinfo {author} {\bibfnamefont {D.~L.}\ \bibnamefont {Bruhwiler}}, \bibinfo {author} {\bibfnamefont {G.}~\bibnamefont {Pretzler}}, \bibinfo {author} {\bibfnamefont {T.}~\bibnamefont {Wilson}}, \bibinfo {author} {\bibfnamefont {Z.}~\bibnamefont {Sheng}}, \bibinfo {author} {\bibfnamefont {M.}~\bibnamefont {Stumpf}}, \bibinfo {author} {\bibfnamefont {L.}~\bibnamefont {Reichwein}}, \bibinfo {author} {\bibfnamefont {A.}~\bibnamefont {Pukhov}}, \bibinfo {author} {\bibfnamefont {J.~R.}\ \bibnamefont {Cary}}, \bibinfo {author} {\bibfnamefont {M.~J.}\ \bibnamefont {Hogan}}, \bibinfo {author} {\bibfnamefont {V.}~\bibnamefont {Yakimenko}}, \bibinfo {author} {\bibfnamefont {J.~B.}\ \bibnamefont {Rosenzweig}},\ and\ \bibinfo {author} {\bibfnamefont {B.}~\bibnamefont {Hidding}},\ }\bibfield  {title} {\bibinfo {title} {{Plasma Photocathodes}},\ }\href
  {https://doi.org/https://doi.org/10.1002/andp.202200655} {\bibfield  {journal} {\bibinfo  {journal} {Annalen der Physik}\ }\textbf {\bibinfo {volume} {535}},\ \bibinfo {pages} {2200655} (\bibinfo {year} {2023}{\natexlab{a}})}\BibitemShut {NoStop}%
\bibitem [{\citenamefont {{Di Mitri}}\ and\ \citenamefont {Cornacchia}(2014)}]{dimitri2014}%
  \BibitemOpen
  \bibfield  {author} {\bibinfo {author} {\bibfnamefont {S.}~\bibnamefont {{Di Mitri}}}\ and\ \bibinfo {author} {\bibfnamefont {M.}~\bibnamefont {Cornacchia}},\ }\bibfield  {title} {\bibinfo {title} {Electron beam brightness in linac drivers for free-electron-lasers},\ }\href {https://doi.org/https://doi.org/10.1016/j.physrep.2014.01.005} {\bibfield  {journal} {\bibinfo  {journal} {Physics Reports}\ }\textbf {\bibinfo {volume} {539}},\ \bibinfo {pages} {1} (\bibinfo {year} {2014})}\BibitemShut {NoStop}%
\bibitem [{\citenamefont {Tajima}\ and\ \citenamefont {Dawson}(1979)}]{TajimaDawsonPRL1979}%
  \BibitemOpen
  \bibfield  {author} {\bibinfo {author} {\bibfnamefont {T.}~\bibnamefont {Tajima}}\ and\ \bibinfo {author} {\bibfnamefont {J.~M.}\ \bibnamefont {Dawson}},\ }\bibfield  {title} {\bibinfo {title} {{Laser Electron Accelerator}},\ }\href {https://doi.org/10.1103/PhysRevLett.43.267} {\bibfield  {journal} {\bibinfo  {journal} {Phys. Rev. Lett.}\ }\textbf {\bibinfo {volume} {43}},\ \bibinfo {pages} {267} (\bibinfo {year} {1979})}\BibitemShut {NoStop}%
\bibitem [{\citenamefont {Ruth}\ \emph {et~al.}(1984)\citenamefont {Ruth}, \citenamefont {Chao}, \citenamefont {Morton},\ and\ \citenamefont {Wilson}}]{Ruth1984}%
  \BibitemOpen
  \bibfield  {author} {\bibinfo {author} {\bibfnamefont {R.~D.}\ \bibnamefont {Ruth}}, \bibinfo {author} {\bibfnamefont {A.~W.}\ \bibnamefont {Chao}}, \bibinfo {author} {\bibfnamefont {P.~L.}\ \bibnamefont {Morton}},\ and\ \bibinfo {author} {\bibfnamefont {P.~B.}\ \bibnamefont {Wilson}},\ }\href {https://doi.org/10.2172/1447278} {\emph {\bibinfo {title} {A Plasma Wake Field Accelerator}}},\ \bibinfo {type} {Tech. Rep.}\ (\bibinfo  {institution} {SLAC National Accelerator Laboratory (SLAC), Menlo Park, CA (United States)},\ \bibinfo {year} {1984})\BibitemShut {NoStop}%
\bibitem [{\citenamefont {Chen}\ \emph {et~al.}(1985)\citenamefont {Chen}, \citenamefont {Dawson}, \citenamefont {Huff},\ and\ \citenamefont {Katsouleas}}]{ChenPRL1985}%
  \BibitemOpen
  \bibfield  {author} {\bibinfo {author} {\bibfnamefont {P.}~\bibnamefont {Chen}}, \bibinfo {author} {\bibfnamefont {J.~M.}\ \bibnamefont {Dawson}}, \bibinfo {author} {\bibfnamefont {R.~W.}\ \bibnamefont {Huff}},\ and\ \bibinfo {author} {\bibfnamefont {T.}~\bibnamefont {Katsouleas}},\ }\bibfield  {title} {\bibinfo {title} {{Acceleration of Electrons by the Interaction of a Bunched Electron Beam with a Plasma}},\ }\href {https://doi.org/10.1103/PhysRevLett.54.693} {\bibfield  {journal} {\bibinfo  {journal} {Phys. Rev. Lett.}\ }\textbf {\bibinfo {volume} {54}},\ \bibinfo {pages} {693} (\bibinfo {year} {1985})}\BibitemShut {NoStop}%
\bibitem [{\citenamefont {Chen}\ \emph {et~al.}(2001)\citenamefont {Chen}, \citenamefont {Cheshkov}, \citenamefont {Ruth},\ and\ \citenamefont {Tajima}}]{Chen2001}%
  \BibitemOpen
  \bibfield  {author} {\bibinfo {author} {\bibfnamefont {P.}~\bibnamefont {Chen}}, \bibinfo {author} {\bibfnamefont {S.}~\bibnamefont {Cheshkov}}, \bibinfo {author} {\bibfnamefont {R.}~\bibnamefont {Ruth}},\ and\ \bibinfo {author} {\bibfnamefont {T.}~\bibnamefont {Tajima}},\ }\bibfield  {title} {\bibinfo {title} {An ultra-high gradient plasma wakefield booster},\ }\href {https://doi.org/10.1063/1.1384416} {\bibfield  {journal} {\bibinfo  {journal} {AIP Conference Proceedings}\ }\textbf {\bibinfo {volume} {569}},\ \bibinfo {pages} {903} (\bibinfo {year} {2001})}\BibitemShut {NoStop}%
\bibitem [{\citenamefont {Blumenfeld}\ \emph {et~al.}(2007)\citenamefont {Blumenfeld}, \citenamefont {Clayton}, \citenamefont {Decker}, \citenamefont {Hogan}, \citenamefont {Huang}, \citenamefont {Ischebeck}, \citenamefont {Iverson}, \citenamefont {Joshi}, \citenamefont {Katsouleas}, \citenamefont {Kirby}, \citenamefont {Lu}, \citenamefont {Marsh}, \citenamefont {Mori}, \citenamefont {Muggli}, \citenamefont {Oz}, \citenamefont {Siemann}, \citenamefont {Walz},\ and\ \citenamefont {Zhou}}]{Blumenfeld2007}%
  \BibitemOpen
  \bibfield  {author} {\bibinfo {author} {\bibfnamefont {I.}~\bibnamefont {Blumenfeld}}, \bibinfo {author} {\bibfnamefont {C.~E.}\ \bibnamefont {Clayton}}, \bibinfo {author} {\bibfnamefont {F.-J.}\ \bibnamefont {Decker}}, \bibinfo {author} {\bibfnamefont {M.~J.}\ \bibnamefont {Hogan}}, \bibinfo {author} {\bibfnamefont {C.}~\bibnamefont {Huang}}, \bibinfo {author} {\bibfnamefont {R.}~\bibnamefont {Ischebeck}}, \bibinfo {author} {\bibfnamefont {R.}~\bibnamefont {Iverson}}, \bibinfo {author} {\bibfnamefont {C.}~\bibnamefont {Joshi}}, \bibinfo {author} {\bibfnamefont {T.}~\bibnamefont {Katsouleas}}, \bibinfo {author} {\bibfnamefont {N.}~\bibnamefont {Kirby}}, \bibinfo {author} {\bibfnamefont {W.}~\bibnamefont {Lu}}, \bibinfo {author} {\bibfnamefont {K.~A.}\ \bibnamefont {Marsh}}, \bibinfo {author} {\bibfnamefont {W.~B.}\ \bibnamefont {Mori}}, \bibinfo {author} {\bibfnamefont {P.}~\bibnamefont {Muggli}}, \bibinfo {author} {\bibfnamefont {E.}~\bibnamefont {Oz}}, \bibinfo {author} {\bibfnamefont {R.~H.}\ \bibnamefont {Siemann}}, \bibinfo {author} {\bibfnamefont {D.}~\bibnamefont {Walz}},\ and\ \bibinfo {author} {\bibfnamefont {M.}~\bibnamefont {Zhou}},\ }\bibfield  {title} {\bibinfo {title} {Energy doubling of 42 {GeV} electrons in a metre-scale plasma wakefield accelerator},\ }\href@noop {} {\bibfield  {journal} {\bibinfo  {journal} {Nature}\ }\textbf {\bibinfo {volume} {445}},\ \bibinfo {pages} {741} (\bibinfo {year} {2007})}\BibitemShut {NoStop}%
\bibitem [{\citenamefont {Hidding}\ \emph {et~al.}(2012)\citenamefont {Hidding}, \citenamefont {Pretzler}, \citenamefont {Rosenzweig}, \citenamefont {K\"onigstein}, \citenamefont {Schiller},\ and\ \citenamefont {Bruhwiler}}]{Hidding2012PRL}%
  \BibitemOpen
  \bibfield  {author} {\bibinfo {author} {\bibfnamefont {B.}~\bibnamefont {Hidding}}, \bibinfo {author} {\bibfnamefont {G.}~\bibnamefont {Pretzler}}, \bibinfo {author} {\bibfnamefont {J.~B.}\ \bibnamefont {Rosenzweig}}, \bibinfo {author} {\bibfnamefont {T.}~\bibnamefont {K\"onigstein}}, \bibinfo {author} {\bibfnamefont {D.}~\bibnamefont {Schiller}},\ and\ \bibinfo {author} {\bibfnamefont {D.~L.}\ \bibnamefont {Bruhwiler}},\ }\bibfield  {title} {\bibinfo {title} {{Ultracold Electron Bunch Generation via Plasma Photocathode Emission and Acceleration in a Beam-Driven Plasma Blowout}},\ }\href {https://doi.org/10.1103/PhysRevLett.108.035001} {\bibfield  {journal} {\bibinfo  {journal} {Phys. Rev. Lett.}\ }\textbf {\bibinfo {volume} {108}},\ \bibinfo {pages} {035001} (\bibinfo {year} {2012})}\BibitemShut {NoStop}%
\bibitem [{\citenamefont {Schroeder}\ \emph {et~al.}(2014)\citenamefont {Schroeder}, \citenamefont {Vay}, \citenamefont {Esarey}, \citenamefont {Bulanov}, \citenamefont {Benedetti}, \citenamefont {Yu}, \citenamefont {Chen}, \citenamefont {Geddes},\ and\ \citenamefont {Leemans}}]{Schroeder2014}%
  \BibitemOpen
  \bibfield  {author} {\bibinfo {author} {\bibfnamefont {C.~B.}\ \bibnamefont {Schroeder}}, \bibinfo {author} {\bibfnamefont {J.-L.}\ \bibnamefont {Vay}}, \bibinfo {author} {\bibfnamefont {E.}~\bibnamefont {Esarey}}, \bibinfo {author} {\bibfnamefont {S.~S.}\ \bibnamefont {Bulanov}}, \bibinfo {author} {\bibfnamefont {C.}~\bibnamefont {Benedetti}}, \bibinfo {author} {\bibfnamefont {L.-L.}\ \bibnamefont {Yu}}, \bibinfo {author} {\bibfnamefont {M.}~\bibnamefont {Chen}}, \bibinfo {author} {\bibfnamefont {C.~G.~R.}\ \bibnamefont {Geddes}},\ and\ \bibinfo {author} {\bibfnamefont {W.~P.}\ \bibnamefont {Leemans}},\ }\bibfield  {title} {\bibinfo {title} {{Thermal emittance from ionization-induced trapping in plasma accelerators}},\ }\href {https://doi.org/10.1103/PhysRevSTAB.17.101301} {\bibfield  {journal} {\bibinfo  {journal} {Phys. Rev. ST Accel. Beams}\ }\textbf {\bibinfo {volume} {17}},\ \bibinfo {pages} {101301} (\bibinfo {year} {2014})}\BibitemShut {NoStop}%
\bibitem [{\citenamefont {Hidding}\ \emph {et~al.}(2014)\citenamefont {Hidding}, \citenamefont {Manahan}, \citenamefont {Karger}, \citenamefont {Knetsch}, \citenamefont {Wittig}, \citenamefont {Jaroszynski}, \citenamefont {Sheng}, \citenamefont {Xi}, \citenamefont {Deng}, \citenamefont {Rosenzweig}, \citenamefont {Andonian}, \citenamefont {Murokh}, \citenamefont {Pretzler}, \citenamefont {Bruhwiler},\ and\ \citenamefont {Smith}}]{Hidding2014}%
  \BibitemOpen
  \bibfield  {author} {\bibinfo {author} {\bibfnamefont {B.}~\bibnamefont {Hidding}}, \bibinfo {author} {\bibfnamefont {G.~G.}\ \bibnamefont {Manahan}}, \bibinfo {author} {\bibfnamefont {O.}~\bibnamefont {Karger}}, \bibinfo {author} {\bibfnamefont {A.}~\bibnamefont {Knetsch}}, \bibinfo {author} {\bibfnamefont {G.}~\bibnamefont {Wittig}}, \bibinfo {author} {\bibfnamefont {D.~A.}\ \bibnamefont {Jaroszynski}}, \bibinfo {author} {\bibfnamefont {Z.-M.}\ \bibnamefont {Sheng}}, \bibinfo {author} {\bibfnamefont {Y.}~\bibnamefont {Xi}}, \bibinfo {author} {\bibfnamefont {A.}~\bibnamefont {Deng}}, \bibinfo {author} {\bibfnamefont {J.~B.}\ \bibnamefont {Rosenzweig}}, \bibinfo {author} {\bibfnamefont {G.}~\bibnamefont {Andonian}}, \bibinfo {author} {\bibfnamefont {A.}~\bibnamefont {Murokh}}, \bibinfo {author} {\bibfnamefont {G.}~\bibnamefont {Pretzler}}, \bibinfo {author} {\bibfnamefont {D.~L.}\ \bibnamefont {Bruhwiler}},\ and\ \bibinfo {author} {\bibfnamefont {J.}~\bibnamefont {Smith}},\ }\bibfield  {title} {\bibinfo {title} {{Ultrahigh brightness bunches from hybrid plasma accelerators as drivers of 5th generation light sources}},\ }\href {https://doi.org/10.1088/0953-4075/47/23/234010} {\bibfield  {journal} {\bibinfo  {journal} {J. Phys. B: At. Mol. Opt. Phys.}\ }\textbf {\bibinfo {volume} {47}},\ \bibinfo {pages} {234010} (\bibinfo {year} {2014})}\BibitemShut {NoStop}%
\bibitem [{\citenamefont {Habib}\ \emph {et~al.}(2023{\natexlab{b}})\citenamefont {Habib}, \citenamefont {Manahan}, \citenamefont {Scherkl}, \citenamefont {Heinemann}, \citenamefont {Sutherland}, \citenamefont {Altuiri}, \citenamefont {Alotaibi}, \citenamefont {Litos}, \citenamefont {Cary}, \citenamefont {Raubenheimer}, \citenamefont {Hemsing}, \citenamefont {Hogan}, \citenamefont {Rosenzweig}, \citenamefont {Williams}, \citenamefont {McNeil},\ and\ \citenamefont {Hidding}}]{Habib2023}%
  \BibitemOpen
  \bibfield  {author} {\bibinfo {author} {\bibfnamefont {A.~F.}\ \bibnamefont {Habib}}, \bibinfo {author} {\bibfnamefont {G.~G.}\ \bibnamefont {Manahan}}, \bibinfo {author} {\bibfnamefont {P.}~\bibnamefont {Scherkl}}, \bibinfo {author} {\bibfnamefont {T.}~\bibnamefont {Heinemann}}, \bibinfo {author} {\bibfnamefont {A.}~\bibnamefont {Sutherland}}, \bibinfo {author} {\bibfnamefont {R.}~\bibnamefont {Altuiri}}, \bibinfo {author} {\bibfnamefont {B.~M.}\ \bibnamefont {Alotaibi}}, \bibinfo {author} {\bibfnamefont {M.}~\bibnamefont {Litos}}, \bibinfo {author} {\bibfnamefont {J.}~\bibnamefont {Cary}}, \bibinfo {author} {\bibfnamefont {T.}~\bibnamefont {Raubenheimer}}, \bibinfo {author} {\bibfnamefont {E.}~\bibnamefont {Hemsing}}, \bibinfo {author} {\bibfnamefont {M.~J.}\ \bibnamefont {Hogan}}, \bibinfo {author} {\bibfnamefont {J.~B.}\ \bibnamefont {Rosenzweig}}, \bibinfo {author} {\bibfnamefont {P.~H.}\ \bibnamefont {Williams}}, \bibinfo {author} {\bibfnamefont {B.~W.~J.}\ \bibnamefont {McNeil}},\ and\ \bibinfo {author} {\bibfnamefont {B.}~\bibnamefont {Hidding}},\ }\bibfield  {title} {\bibinfo {title} {{Attosecond-Angstrom free-electron-laser towards the cold beam limit}},\ }\href {https://doi.org/10.1038/s41467-023-36592-z} {\bibfield  {journal} {\bibinfo  {journal} {Nat. Commun.}\ }\textbf {\bibinfo {volume} {14}},\ \bibinfo {pages} {1054} (\bibinfo {year} {2023}{\natexlab{b}})}\BibitemShut {NoStop}%
\bibitem [{\citenamefont {Tzoufras}\ \emph {et~al.}(2008)\citenamefont {Tzoufras}, \citenamefont {Lu}, \citenamefont {Tsung}, \citenamefont {Huang}, \citenamefont {Mori}, \citenamefont {Katsouleas}, \citenamefont {Vieira}, \citenamefont {Fonseca},\ and\ \citenamefont {Silva}}]{Tzoufras2008}%
  \BibitemOpen
  \bibfield  {author} {\bibinfo {author} {\bibfnamefont {M.}~\bibnamefont {Tzoufras}}, \bibinfo {author} {\bibfnamefont {W.}~\bibnamefont {Lu}}, \bibinfo {author} {\bibfnamefont {F.~S.}\ \bibnamefont {Tsung}}, \bibinfo {author} {\bibfnamefont {C.}~\bibnamefont {Huang}}, \bibinfo {author} {\bibfnamefont {W.~B.}\ \bibnamefont {Mori}}, \bibinfo {author} {\bibfnamefont {T.}~\bibnamefont {Katsouleas}}, \bibinfo {author} {\bibfnamefont {J.}~\bibnamefont {Vieira}}, \bibinfo {author} {\bibfnamefont {R.~A.}\ \bibnamefont {Fonseca}},\ and\ \bibinfo {author} {\bibfnamefont {L.~O.}\ \bibnamefont {Silva}},\ }\bibfield  {title} {\bibinfo {title} {{Beam Loading in the Nonlinear Regime of Plasma-Based Acceleration}},\ }\href {https://doi.org/10.1103/PhysRevLett.101.145002} {\bibfield  {journal} {\bibinfo  {journal} {Phys. Rev. Lett.}\ }\textbf {\bibinfo {volume} {101}},\ \bibinfo {pages} {145002} (\bibinfo {year} {2008})}\BibitemShut {NoStop}%
\bibitem [{\citenamefont {Diederichs}\ \emph {et~al.}(2024)\citenamefont {Diederichs}, \citenamefont {Benedetti}, \citenamefont {Pousa}, \citenamefont {Sinn}, \citenamefont {Osterhoff}, \citenamefont {Schroeder},\ and\ \citenamefont {Thévenet}}]{diederichs2024}%
  \BibitemOpen
  \bibfield  {author} {\bibinfo {author} {\bibfnamefont {S.}~\bibnamefont {Diederichs}}, \bibinfo {author} {\bibfnamefont {C.}~\bibnamefont {Benedetti}}, \bibinfo {author} {\bibfnamefont {A.~F.}\ \bibnamefont {Pousa}}, \bibinfo {author} {\bibfnamefont {A.}~\bibnamefont {Sinn}}, \bibinfo {author} {\bibfnamefont {J.}~\bibnamefont {Osterhoff}}, \bibinfo {author} {\bibfnamefont {C.~B.}\ \bibnamefont {Schroeder}},\ and\ \bibinfo {author} {\bibfnamefont {M.}~\bibnamefont {Thévenet}},\ }\href {https://arxiv.org/abs/2403.05871} {\bibinfo {title} {Resonant emittance mixing of flat beams in plasma accelerators}} (\bibinfo {year} {2024}),\ \Eprint {https://arxiv.org/abs/2403.05871} {arXiv:2403.05871 [physics.acc-ph]} \BibitemShut {NoStop}%
\bibitem [{\citenamefont {Jolly}\ and\ \citenamefont {Porras}(2022)}]{Jolly2022OptLett}%
  \BibitemOpen
  \bibfield  {author} {\bibinfo {author} {\bibfnamefont {S.~W.}\ \bibnamefont {Jolly}}\ and\ \bibinfo {author} {\bibfnamefont {M.~A.}\ \bibnamefont {Porras}},\ }\bibfield  {title} {\bibinfo {title} {{Clarification for the fields of different radially polarized Laguerre--Gaussian light beams}},\ }\href {https://doi.org/10.1364/OL.464118} {\bibfield  {journal} {\bibinfo  {journal} {Opt. Lett.}\ }\textbf {\bibinfo {volume} {47}},\ \bibinfo {pages} {3632} (\bibinfo {year} {2022})}\BibitemShut {NoStop}%
\bibitem [{\citenamefont {Thévenet}\ \emph {et~al.}(2024)\citenamefont {Thévenet}, \citenamefont {Andriyash}, \citenamefont {Fedeli}, \citenamefont {Ángel Ferran~Pousa}, \citenamefont {Huebl}, \citenamefont {Jalas}, \citenamefont {Kirchen}, \citenamefont {Lehe}, \citenamefont {Shalloo}, \citenamefont {Sinn},\ and\ \citenamefont {Vay}}]{LASY}%
  \BibitemOpen
  \bibfield  {author} {\bibinfo {author} {\bibfnamefont {M.}~\bibnamefont {Thévenet}}, \bibinfo {author} {\bibfnamefont {I.~A.}\ \bibnamefont {Andriyash}}, \bibinfo {author} {\bibfnamefont {L.}~\bibnamefont {Fedeli}}, \bibinfo {author} {\bibnamefont {Ángel Ferran~Pousa}}, \bibinfo {author} {\bibfnamefont {A.}~\bibnamefont {Huebl}}, \bibinfo {author} {\bibfnamefont {S.}~\bibnamefont {Jalas}}, \bibinfo {author} {\bibfnamefont {M.}~\bibnamefont {Kirchen}}, \bibinfo {author} {\bibfnamefont {R.}~\bibnamefont {Lehe}}, \bibinfo {author} {\bibfnamefont {R.~J.}\ \bibnamefont {Shalloo}}, \bibinfo {author} {\bibfnamefont {A.}~\bibnamefont {Sinn}},\ and\ \bibinfo {author} {\bibfnamefont {J.-L.}\ \bibnamefont {Vay}},\ }\href {https://arxiv.org/abs/2403.12191} {\bibinfo {title} {{LASY: LAser manipulations made eaSY}}} (\bibinfo {year} {2024}),\ \Eprint {https://arxiv.org/abs/2403.12191} {arXiv:2403.12191 [physics.optics]} \BibitemShut {NoStop}%
\bibitem [{\citenamefont {Lehe}\ \emph {et~al.}(2016)\citenamefont {Lehe}, \citenamefont {Kirchen}, \citenamefont {Andriyash}, \citenamefont {Godfrey},\ and\ \citenamefont {Vay}}]{FBPICLehe2016}%
  \BibitemOpen
  \bibfield  {author} {\bibinfo {author} {\bibfnamefont {R.}~\bibnamefont {Lehe}}, \bibinfo {author} {\bibfnamefont {M.}~\bibnamefont {Kirchen}}, \bibinfo {author} {\bibfnamefont {I.~A.}\ \bibnamefont {Andriyash}}, \bibinfo {author} {\bibfnamefont {B.~B.}\ \bibnamefont {Godfrey}},\ and\ \bibinfo {author} {\bibfnamefont {J.-L.}\ \bibnamefont {Vay}},\ }\bibfield  {title} {\bibinfo {title} {{A spectral, quasi-cylindrical and dispersion-free Particle-In-Cell algorithm}},\ }\href {https://doi.org/https://doi.org/10.1016/j.cpc.2016.02.007} {\bibfield  {journal} {\bibinfo  {journal} {Computer Physics Communications}\ }\textbf {\bibinfo {volume} {203}},\ \bibinfo {pages} {66} (\bibinfo {year} {2016})}\BibitemShut {NoStop}%
\bibitem [{\citenamefont {Jalas}\ \emph {et~al.}(2017)\citenamefont {Jalas}, \citenamefont {Dornmair}, \citenamefont {Lehe}, \citenamefont {Vincenti}, \citenamefont {Vay}, \citenamefont {Kirchen},\ and\ \citenamefont {Maier}}]{FBPICJalasPoP2017}%
  \BibitemOpen
  \bibfield  {author} {\bibinfo {author} {\bibfnamefont {S.}~\bibnamefont {Jalas}}, \bibinfo {author} {\bibfnamefont {I.}~\bibnamefont {Dornmair}}, \bibinfo {author} {\bibfnamefont {R.}~\bibnamefont {Lehe}}, \bibinfo {author} {\bibfnamefont {H.}~\bibnamefont {Vincenti}}, \bibinfo {author} {\bibfnamefont {J.-L.}\ \bibnamefont {Vay}}, \bibinfo {author} {\bibfnamefont {M.}~\bibnamefont {Kirchen}},\ and\ \bibinfo {author} {\bibfnamefont {A.~R.}\ \bibnamefont {Maier}},\ }\bibfield  {title} {\bibinfo {title} {{Accurate modeling of plasma acceleration with arbitrary order pseudo-spectral particle-in-cell methods}},\ }\href {https://doi.org/10.1063/1.4978569} {\bibfield  {journal} {\bibinfo  {journal} {Physics of Plasmas}\ }\textbf {\bibinfo {volume} {24}},\ \bibinfo {pages} {033115} (\bibinfo {year} {2017})}\BibitemShut {NoStop}%
\bibitem [{\citenamefont {Kirchen}\ \emph {et~al.}(2020)\citenamefont {Kirchen}, \citenamefont {Lehe}, \citenamefont {Jalas}, \citenamefont {Shapoval}, \citenamefont {Vay},\ and\ \citenamefont {Maier}}]{FBPICKirchen2020PRE}%
  \BibitemOpen
  \bibfield  {author} {\bibinfo {author} {\bibfnamefont {M.}~\bibnamefont {Kirchen}}, \bibinfo {author} {\bibfnamefont {R.}~\bibnamefont {Lehe}}, \bibinfo {author} {\bibfnamefont {S.}~\bibnamefont {Jalas}}, \bibinfo {author} {\bibfnamefont {O.}~\bibnamefont {Shapoval}}, \bibinfo {author} {\bibfnamefont {J.-L.}\ \bibnamefont {Vay}},\ and\ \bibinfo {author} {\bibfnamefont {A.~R.}\ \bibnamefont {Maier}},\ }\bibfield  {title} {\bibinfo {title} {{Scalable spectral solver in Galilean coordinates for eliminating the numerical Cherenkov instability in particle-in-cell simulations of streaming plasmas}},\ }\href {https://doi.org/10.1103/PhysRevE.102.013202} {\bibfield  {journal} {\bibinfo  {journal} {Phys. Rev. E}\ }\textbf {\bibinfo {volume} {102}},\ \bibinfo {pages} {013202} (\bibinfo {year} {2020})}\BibitemShut {NoStop}%
\bibitem [{\citenamefont {Ferran~Pousa}\ \emph {et~al.}(2019)\citenamefont {Ferran~Pousa}, \citenamefont {Assmann},\ and\ \citenamefont {Martinez de~la Ossa}}]{Wake-TFerranPousa2019}%
  \BibitemOpen
  \bibfield  {author} {\bibinfo {author} {\bibfnamefont {A.}~\bibnamefont {Ferran~Pousa}}, \bibinfo {author} {\bibfnamefont {R.}~\bibnamefont {Assmann}},\ and\ \bibinfo {author} {\bibfnamefont {A.}~\bibnamefont {Martinez de~la Ossa}},\ }\bibfield  {title} {\bibinfo {title} {Wake-{T}: a fast particle tracking code for plasma-based accelerators},\ }\href {https://doi.org/10.1088/1742-6596/1350/1/012056} {\bibfield  {journal} {\bibinfo  {journal} {Journal of Physics: Conference Series}\ }\textbf {\bibinfo {volume} {1350}},\ \bibinfo {pages} {012056} (\bibinfo {year} {2019})}\BibitemShut {NoStop}%
\bibitem [{\citenamefont {D'Arcy}\ \emph {et~al.}(2019)\citenamefont {D'Arcy}, \citenamefont {Aschikhin}, \citenamefont {Bohlen}, \citenamefont {Boyle}, \citenamefont {Br{\"u}mmer}, \citenamefont {Chappell}, \citenamefont {Diederichs}, \citenamefont {Foster}, \citenamefont {Garland}, \citenamefont {Goldberg}, \citenamefont {Gonzalez}, \citenamefont {Karstensen}, \citenamefont {Knetsch}, \citenamefont {Kuang}, \citenamefont {Libov}, \citenamefont {Ludwig}, \citenamefont {Martinez de~la Ossa}, \citenamefont {Marutzky}, \citenamefont {Meisel}, \citenamefont {Mehrling}, \citenamefont {Niknejadi}, \citenamefont {P{\~o}der}, \citenamefont {Pourmoussavi}, \citenamefont {Quast}, \citenamefont {R{\"o}ckemann}, \citenamefont {Schaper}, \citenamefont {Schmidt}, \citenamefont {Schr{\"o}der}, \citenamefont {Schwinkendorf}, \citenamefont {Sheeran}, \citenamefont {Tauscher}, \citenamefont {Wesch}, \citenamefont {Wing}, \citenamefont {Winkler}, \citenamefont {Zeng},\ and\ \citenamefont {Osterhoff}}]{FLASHForwardDArcy2019}%
  \BibitemOpen
  \bibfield  {author} {\bibinfo {author} {\bibfnamefont {R.}~\bibnamefont {D'Arcy}}, \bibinfo {author} {\bibfnamefont {A.}~\bibnamefont {Aschikhin}}, \bibinfo {author} {\bibfnamefont {S.}~\bibnamefont {Bohlen}}, \bibinfo {author} {\bibfnamefont {G.}~\bibnamefont {Boyle}}, \bibinfo {author} {\bibfnamefont {T.}~\bibnamefont {Br{\"u}mmer}}, \bibinfo {author} {\bibfnamefont {J.}~\bibnamefont {Chappell}}, \bibinfo {author} {\bibfnamefont {S.}~\bibnamefont {Diederichs}}, \bibinfo {author} {\bibfnamefont {B.}~\bibnamefont {Foster}}, \bibinfo {author} {\bibfnamefont {M.~J.}\ \bibnamefont {Garland}}, \bibinfo {author} {\bibfnamefont {L.}~\bibnamefont {Goldberg}}, \bibinfo {author} {\bibfnamefont {P.}~\bibnamefont {Gonzalez}}, \bibinfo {author} {\bibfnamefont {S.}~\bibnamefont {Karstensen}}, \bibinfo {author} {\bibfnamefont {A.}~\bibnamefont {Knetsch}}, \bibinfo {author} {\bibfnamefont {P.}~\bibnamefont {Kuang}}, \bibinfo {author} {\bibfnamefont {V.}~\bibnamefont {Libov}}, \bibinfo {author} {\bibfnamefont {K.}~\bibnamefont {Ludwig}}, \bibinfo {author} {\bibfnamefont {A.}~\bibnamefont {Martinez de~la Ossa}}, \bibinfo {author} {\bibfnamefont {F.}~\bibnamefont {Marutzky}}, \bibinfo {author} {\bibfnamefont {M.}~\bibnamefont {Meisel}}, \bibinfo {author} {\bibfnamefont {T.~J.}\ \bibnamefont {Mehrling}}, \bibinfo {author} {\bibfnamefont {P.}~\bibnamefont {Niknejadi}}, \bibinfo {author} {\bibfnamefont {K.}~\bibnamefont {P{\~o}der}}, \bibinfo {author} {\bibfnamefont {P.}~\bibnamefont {Pourmoussavi}}, \bibinfo {author} {\bibfnamefont {M.}~\bibnamefont {Quast}}, \bibinfo {author} {\bibfnamefont {J.-H.}\ \bibnamefont {R{\"o}ckemann}}, \bibinfo {author} {\bibfnamefont {L.}~\bibnamefont {Schaper}}, \bibinfo {author} {\bibfnamefont {B.}~\bibnamefont {Schmidt}}, \bibinfo {author} {\bibfnamefont {S.}~\bibnamefont {Schr{\"o}der}}, \bibinfo {author} {\bibfnamefont {J.-P.}\ \bibnamefont {Schwinkendorf}}, \bibinfo {author} {\bibfnamefont {B.}~\bibnamefont {Sheeran}}, \bibinfo {author} {\bibfnamefont {G.}~\bibnamefont {Tauscher}}, \bibinfo
  {author} {\bibfnamefont {S.}~\bibnamefont {Wesch}}, \bibinfo {author} {\bibfnamefont {M.}~\bibnamefont {Wing}}, \bibinfo {author} {\bibfnamefont {P.}~\bibnamefont {Winkler}}, \bibinfo {author} {\bibfnamefont {M.}~\bibnamefont {Zeng}},\ and\ \bibinfo {author} {\bibfnamefont {J.}~\bibnamefont {Osterhoff}},\ }\bibfield  {title} {\bibinfo {title} {{FLASHForward}: plasma wakefield accelerator science for high-average-power applications},\ }\href@noop {} {\bibfield  {journal} {\bibinfo  {journal} {Phil. Trans. R. Soc. A}\ }\textbf {\bibinfo {volume} {377}},\ \bibinfo {pages} {20180392} (\bibinfo {year} {2019})}\BibitemShut {NoStop}%
\bibitem [{\citenamefont {Yakimenko}\ \emph {et~al.}(2019)\citenamefont {Yakimenko}, \citenamefont {Alsberg}, \citenamefont {Bong}, \citenamefont {Bouchard}, \citenamefont {Clarke}, \citenamefont {Emma}, \citenamefont {Green}, \citenamefont {Hast}, \citenamefont {Hogan}, \citenamefont {Seabury}, \citenamefont {Lipkowitz}, \citenamefont {O'Shea}, \citenamefont {Storey}, \citenamefont {White},\ and\ \citenamefont {Yocky}}]{FACET-II}%
  \BibitemOpen
  \bibfield  {author} {\bibinfo {author} {\bibfnamefont {V.}~\bibnamefont {Yakimenko}}, \bibinfo {author} {\bibfnamefont {L.}~\bibnamefont {Alsberg}}, \bibinfo {author} {\bibfnamefont {E.}~\bibnamefont {Bong}}, \bibinfo {author} {\bibfnamefont {G.}~\bibnamefont {Bouchard}}, \bibinfo {author} {\bibfnamefont {C.}~\bibnamefont {Clarke}}, \bibinfo {author} {\bibfnamefont {C.}~\bibnamefont {Emma}}, \bibinfo {author} {\bibfnamefont {S.}~\bibnamefont {Green}}, \bibinfo {author} {\bibfnamefont {C.}~\bibnamefont {Hast}}, \bibinfo {author} {\bibfnamefont {M.~J.}\ \bibnamefont {Hogan}}, \bibinfo {author} {\bibfnamefont {J.}~\bibnamefont {Seabury}}, \bibinfo {author} {\bibfnamefont {N.}~\bibnamefont {Lipkowitz}}, \bibinfo {author} {\bibfnamefont {B.}~\bibnamefont {O'Shea}}, \bibinfo {author} {\bibfnamefont {D.}~\bibnamefont {Storey}}, \bibinfo {author} {\bibfnamefont {G.}~\bibnamefont {White}},\ and\ \bibinfo {author} {\bibfnamefont {G.}~\bibnamefont {Yocky}},\ }\bibfield  {title} {\bibinfo {title} {{FACET-II facility for advanced accelerator experimental tests}},\ }\href {https://doi.org/10.1103/PhysRevAccelBeams.22.101301} {\bibfield  {journal} {\bibinfo  {journal} {Phys. Rev. Accel. Beams}\ }\textbf {\bibinfo {volume} {22}},\ \bibinfo {pages} {101301} (\bibinfo {year} {2019})}\BibitemShut {NoStop}%
\bibitem [{\citenamefont {Kurz}\ \emph {et~al.}(2021)\citenamefont {Kurz}, \citenamefont {Heinemann}, \citenamefont {Gilljohann}, \citenamefont {Chang}, \citenamefont {Couperus~Cabada{\u{g}}}, \citenamefont {Debus}, \citenamefont {Kononenko}, \citenamefont {Pausch}, \citenamefont {Sch{\"o}bel}, \citenamefont {Assmann}, \citenamefont {Bussmann}, \citenamefont {Ding}, \citenamefont {G{\"o}tzfried}, \citenamefont {K{\"o}hler}, \citenamefont {Raj}, \citenamefont {Schindler}, \citenamefont {Steiniger}, \citenamefont {Zarini}, \citenamefont {Corde}, \citenamefont {D{\"o}pp}, \citenamefont {Hidding}, \citenamefont {Karsch}, \citenamefont {Schramm}, \citenamefont {Martinez de~la Ossa},\ and\ \citenamefont {Irman}}]{Kurz2021}%
  \BibitemOpen
  \bibfield  {author} {\bibinfo {author} {\bibfnamefont {T.}~\bibnamefont {Kurz}}, \bibinfo {author} {\bibfnamefont {T.}~\bibnamefont {Heinemann}}, \bibinfo {author} {\bibfnamefont {M.~F.}\ \bibnamefont {Gilljohann}}, \bibinfo {author} {\bibfnamefont {Y.~Y.}\ \bibnamefont {Chang}}, \bibinfo {author} {\bibfnamefont {J.~P.}\ \bibnamefont {Couperus~Cabada{\u{g}}}}, \bibinfo {author} {\bibfnamefont {A.}~\bibnamefont {Debus}}, \bibinfo {author} {\bibfnamefont {O.}~\bibnamefont {Kononenko}}, \bibinfo {author} {\bibfnamefont {R.}~\bibnamefont {Pausch}}, \bibinfo {author} {\bibfnamefont {S.}~\bibnamefont {Sch{\"o}bel}}, \bibinfo {author} {\bibfnamefont {R.~W.}\ \bibnamefont {Assmann}}, \bibinfo {author} {\bibfnamefont {M.}~\bibnamefont {Bussmann}}, \bibinfo {author} {\bibfnamefont {H.}~\bibnamefont {Ding}}, \bibinfo {author} {\bibfnamefont {J.}~\bibnamefont {G{\"o}tzfried}}, \bibinfo {author} {\bibfnamefont {A.}~\bibnamefont {K{\"o}hler}}, \bibinfo {author} {\bibfnamefont {G.}~\bibnamefont {Raj}}, \bibinfo {author} {\bibfnamefont {S.}~\bibnamefont {Schindler}}, \bibinfo {author} {\bibfnamefont {K.}~\bibnamefont {Steiniger}}, \bibinfo {author} {\bibfnamefont {O.}~\bibnamefont {Zarini}}, \bibinfo {author} {\bibfnamefont {S.}~\bibnamefont {Corde}}, \bibinfo {author} {\bibfnamefont {A.}~\bibnamefont {D{\"o}pp}}, \bibinfo {author} {\bibfnamefont {B.}~\bibnamefont {Hidding}}, \bibinfo {author} {\bibfnamefont {S.}~\bibnamefont {Karsch}}, \bibinfo {author} {\bibfnamefont {U.}~\bibnamefont {Schramm}}, \bibinfo {author} {\bibfnamefont {A.}~\bibnamefont {Martinez de~la Ossa}},\ and\ \bibinfo {author} {\bibfnamefont {A.}~\bibnamefont {Irman}},\ }\bibfield  {title} {\bibinfo {title} {Demonstration of a compact plasma accelerator powered by laser-accelerated electron beams},\ }\href {https://doi.org/10.1038/s41467-021-23000-7} {\bibfield  {journal} {\bibinfo  {journal} {Nat. Commun.}\ }\textbf {\bibinfo {volume} {12}},\ \bibinfo {pages} {2895} (\bibinfo {year} {2021})}\BibitemShut {NoStop}%
\bibitem [{\citenamefont {Foerster}\ \emph {et~al.}(2022)\citenamefont {Foerster}, \citenamefont {D\"opp}, \citenamefont {Haberstroh}, \citenamefont {Grafenstein}, \citenamefont {Campbell}, \citenamefont {Chang}, \citenamefont {Corde}, \citenamefont {Couperus Cabada\ifmmode~\breve{g}\else \u{g}\fi{}}, \citenamefont {Debus}, \citenamefont {Gilljohann}, \citenamefont {Habib}, \citenamefont {Heinemann}, \citenamefont {Hidding}, \citenamefont {Irman}, \citenamefont {Irshad}, \citenamefont {Knetsch}, \citenamefont {Kononenko}, \citenamefont {Martinez de~la Ossa}, \citenamefont {Nutter}, \citenamefont {Pausch}, \citenamefont {Schilling}, \citenamefont {Schletter}, \citenamefont {Sch\"obel}, \citenamefont {Schramm}, \citenamefont {Travac}, \citenamefont {Ufer},\ and\ \citenamefont {Karsch}}]{Foerster2022}%
  \BibitemOpen
  \bibfield  {author} {\bibinfo {author} {\bibfnamefont {F.~M.}\ \bibnamefont {Foerster}}, \bibinfo {author} {\bibfnamefont {A.}~\bibnamefont {D\"opp}}, \bibinfo {author} {\bibfnamefont {F.}~\bibnamefont {Haberstroh}}, \bibinfo {author} {\bibfnamefont {K.~v.}\ \bibnamefont {Grafenstein}}, \bibinfo {author} {\bibfnamefont {D.}~\bibnamefont {Campbell}}, \bibinfo {author} {\bibfnamefont {Y.-Y.}\ \bibnamefont {Chang}}, \bibinfo {author} {\bibfnamefont {S.}~\bibnamefont {Corde}}, \bibinfo {author} {\bibfnamefont {J.~P.}\ \bibnamefont {Couperus Cabada\ifmmode~\breve{g}\else \u{g}\fi{}}}, \bibinfo {author} {\bibfnamefont {A.}~\bibnamefont {Debus}}, \bibinfo {author} {\bibfnamefont {M.~F.}\ \bibnamefont {Gilljohann}}, \bibinfo {author} {\bibfnamefont {A.~F.}\ \bibnamefont {Habib}}, \bibinfo {author} {\bibfnamefont {T.}~\bibnamefont {Heinemann}}, \bibinfo {author} {\bibfnamefont {B.}~\bibnamefont {Hidding}}, \bibinfo {author} {\bibfnamefont {A.}~\bibnamefont {Irman}}, \bibinfo {author} {\bibfnamefont {F.}~\bibnamefont {Irshad}}, \bibinfo {author} {\bibfnamefont {A.}~\bibnamefont {Knetsch}}, \bibinfo {author} {\bibfnamefont {O.}~\bibnamefont {Kononenko}}, \bibinfo {author} {\bibfnamefont {A.}~\bibnamefont {Martinez de~la Ossa}}, \bibinfo {author} {\bibfnamefont {A.}~\bibnamefont {Nutter}}, \bibinfo {author} {\bibfnamefont {R.}~\bibnamefont {Pausch}}, \bibinfo {author} {\bibfnamefont {G.}~\bibnamefont {Schilling}}, \bibinfo {author} {\bibfnamefont {A.}~\bibnamefont {Schletter}}, \bibinfo {author} {\bibfnamefont {S.}~\bibnamefont {Sch\"obel}}, \bibinfo {author} {\bibfnamefont {U.}~\bibnamefont {Schramm}}, \bibinfo {author} {\bibfnamefont {E.}~\bibnamefont {Travac}}, \bibinfo {author} {\bibfnamefont {P.}~\bibnamefont {Ufer}},\ and\ \bibinfo {author} {\bibfnamefont {S.}~\bibnamefont {Karsch}},\ }\bibfield  {title} {\bibinfo {title} {{Stable and High-Quality Electron Beams from Staged Laser and Plasma Wakefield Accelerators}},\ }\href {https://doi.org/10.1103/PhysRevX.12.041016} {\bibfield  {journal} {\bibinfo  {journal} {Phys.
  Rev. X}\ }\textbf {\bibinfo {volume} {12}},\ \bibinfo {pages} {041016} (\bibinfo {year} {2022})}\BibitemShut {NoStop}%
\bibitem [{\citenamefont {Foster}\ \emph {et~al.}(2023)\citenamefont {Foster}, \citenamefont {D’Arcy},\ and\ \citenamefont {Lindstrøm}}]{HALHFFoster2023}%
  \BibitemOpen
  \bibfield  {author} {\bibinfo {author} {\bibfnamefont {B.}~\bibnamefont {Foster}}, \bibinfo {author} {\bibfnamefont {R.}~\bibnamefont {D’Arcy}},\ and\ \bibinfo {author} {\bibfnamefont {C.~A.}\ \bibnamefont {Lindstrøm}},\ }\bibfield  {title} {\bibinfo {title} {A hybrid, asymmetric, linear higgs factory based on plasma-wakefield and radio-frequency acceleration},\ }\href {https://doi.org/10.1088/1367-2630/acf395} {\bibfield  {journal} {\bibinfo  {journal} {New Journal of Physics}\ }\textbf {\bibinfo {volume} {25}},\ \bibinfo {pages} {093037} (\bibinfo {year} {2023})}\BibitemShut {NoStop}%
\bibitem [{\citenamefont {Kawata}\ \emph {et~al.}(2005)\citenamefont {Kawata}, \citenamefont {Kong}, \citenamefont {Miyazaki} \emph {et~al.}}]{KAWATA2005}%
  \BibitemOpen
  \bibfield  {author} {\bibinfo {author} {\bibfnamefont {S.}~\bibnamefont {Kawata}}, \bibinfo {author} {\bibfnamefont {Q.}~\bibnamefont {Kong}}, \bibinfo {author} {\bibfnamefont {S.}~\bibnamefont {Miyazaki}}, \emph {et~al.},\ }\bibfield  {title} {\bibinfo {title} {Electron bunch acceleration and trapping by ponderomotive force of an intense short-pulse laser},\ }\href {https://doi.org/10.1017/S0263034605050123} {\bibfield  {journal} {\bibinfo  {journal} {Laser and Particle Beams}\ }\textbf {\bibinfo {volume} {23}},\ \bibinfo {pages} {61–67} (\bibinfo {year} {2005})}\BibitemShut {NoStop}%
\bibitem [{\citenamefont {Gupta}\ \emph {et~al.}(2007)\citenamefont {Gupta}, \citenamefont {Kant}, \citenamefont {Kim},\ and\ \citenamefont {Suk}}]{GUPTA2007}%
  \BibitemOpen
  \bibfield  {author} {\bibinfo {author} {\bibfnamefont {D.~N.}\ \bibnamefont {Gupta}}, \bibinfo {author} {\bibfnamefont {N.}~\bibnamefont {Kant}}, \bibinfo {author} {\bibfnamefont {D.~E.}\ \bibnamefont {Kim}},\ and\ \bibinfo {author} {\bibfnamefont {H.}~\bibnamefont {Suk}},\ }\bibfield  {title} {\bibinfo {title} {{Electron acceleration to GeV energy by a radially polarized laser}},\ }\href {https://doi.org/https://doi.org/10.1016/j.physleta.2007.04.030} {\bibfield  {journal} {\bibinfo  {journal} {Physics Letters A}\ }\textbf {\bibinfo {volume} {368}},\ \bibinfo {pages} {402} (\bibinfo {year} {2007})}\BibitemShut {NoStop}%
\bibitem [{\citenamefont {Pak}\ \emph {et~al.}(2010)\citenamefont {Pak}, \citenamefont {Marsh}, \citenamefont {Martins}, \citenamefont {Lu}, \citenamefont {Mori},\ and\ \citenamefont {Joshi}}]{Pak2010}%
  \BibitemOpen
  \bibfield  {author} {\bibinfo {author} {\bibfnamefont {A.}~\bibnamefont {Pak}}, \bibinfo {author} {\bibfnamefont {K.~A.}\ \bibnamefont {Marsh}}, \bibinfo {author} {\bibfnamefont {S.~F.}\ \bibnamefont {Martins}}, \bibinfo {author} {\bibfnamefont {W.}~\bibnamefont {Lu}}, \bibinfo {author} {\bibfnamefont {W.~B.}\ \bibnamefont {Mori}},\ and\ \bibinfo {author} {\bibfnamefont {C.}~\bibnamefont {Joshi}},\ }\bibfield  {title} {\bibinfo {title} {{Injection and Trapping of Tunnel-Ionized Electrons into Laser-Produced Wakes}},\ }\href {https://doi.org/10.1103/PhysRevLett.104.025003} {\bibfield  {journal} {\bibinfo  {journal} {Phys. Rev. Lett.}\ }\textbf {\bibinfo {volume} {104}},\ \bibinfo {pages} {025003} (\bibinfo {year} {2010})}\BibitemShut {NoStop}%
\bibitem [{\citenamefont {Emmerich}\ \emph {et~al.}(2006)\citenamefont {Emmerich}, \citenamefont {Giannakoglou},\ and\ \citenamefont {Naujoks}}]{MOBOEmmerich2006}%
  \BibitemOpen
  \bibfield  {author} {\bibinfo {author} {\bibfnamefont {M.}~\bibnamefont {Emmerich}}, \bibinfo {author} {\bibfnamefont {K.}~\bibnamefont {Giannakoglou}},\ and\ \bibinfo {author} {\bibfnamefont {B.}~\bibnamefont {Naujoks}},\ }\bibfield  {title} {\bibinfo {title} {{Single- and multiobjective evolutionary optimization assisted by Gaussian random field metamodels}},\ }\href {https://doi.org/10.1109/TEVC.2005.859463} {\bibfield  {journal} {\bibinfo  {journal} {IEEE Transactions on Evolutionary Computation}\ }\textbf {\bibinfo {volume} {10}},\ \bibinfo {pages} {421} (\bibinfo {year} {2006})}\BibitemShut {NoStop}%
\bibitem [{\citenamefont {Daulton}\ \emph {et~al.}(2020)\citenamefont {Daulton}, \citenamefont {Balandat},\ and\ \citenamefont {Bakshy}}]{MOBODaulton2020}%
  \BibitemOpen
  \bibfield  {author} {\bibinfo {author} {\bibfnamefont {S.}~\bibnamefont {Daulton}}, \bibinfo {author} {\bibfnamefont {M.}~\bibnamefont {Balandat}},\ and\ \bibinfo {author} {\bibfnamefont {E.}~\bibnamefont {Bakshy}},\ }\bibfield  {title} {\bibinfo {title} {{Differentiable expected hypervolume improvement for parallel multi-objective Bayesian optimization}},\ }in\ \href@noop {} {\emph {\bibinfo {booktitle} {Proceedings of the 34th International Conference on Neural Information Processing Systems}}},\ \bibinfo {series and number} {NIPS '20}\ (\bibinfo  {publisher} {Curran Associates Inc.},\ \bibinfo {address} {Red Hook, NY, USA},\ \bibinfo {year} {2020})\BibitemShut {NoStop}%
\bibitem [{\citenamefont {Rasmussen}\ and\ \citenamefont {Williams}(2005)}]{GPRasumussen2005}%
  \BibitemOpen
  \bibfield  {author} {\bibinfo {author} {\bibfnamefont {C.~E.}\ \bibnamefont {Rasmussen}}\ and\ \bibinfo {author} {\bibfnamefont {C.~K.~I.}\ \bibnamefont {Williams}},\ }\href {https://doi.org/10.7551/mitpress/3206.001.0001} {\emph {\bibinfo {title} {{Gaussian Processes for Machine Learning}}}}\ (\bibinfo  {publisher} {The MIT Press},\ \bibinfo {year} {2005})\BibitemShut {NoStop}%
\bibitem [{\citenamefont {Daulton}\ \emph {et~al.}(2021)\citenamefont {Daulton}, \citenamefont {Balandat},\ and\ \citenamefont {Bakshy}}]{MOBODaulton2021}%
  \BibitemOpen
  \bibfield  {author} {\bibinfo {author} {\bibfnamefont {S.}~\bibnamefont {Daulton}}, \bibinfo {author} {\bibfnamefont {M.}~\bibnamefont {Balandat}},\ and\ \bibinfo {author} {\bibfnamefont {E.}~\bibnamefont {Bakshy}},\ }\bibfield  {title} {\bibinfo {title} {Parallel bayesian optimization of multiple noisy objectives with expected hypervolume improvement},\ }in\ \href {https://proceedings.neurips.cc/paper_files/paper/2021/file/11704817e347269b7254e744b5e22dac-Paper.pdf} {\emph {\bibinfo {booktitle} {{Advances in Neural Information Processing Systems}}}},\ Vol.~\bibinfo {volume} {34},\ \bibinfo {editor} {edited by\ \bibinfo {editor} {\bibfnamefont {M.}~\bibnamefont {Ranzato}}, \bibinfo {editor} {\bibfnamefont {A.}~\bibnamefont {Beygelzimer}}, \bibinfo {editor} {\bibfnamefont {Y.}~\bibnamefont {Dauphin}}, \bibinfo {editor} {\bibfnamefont {P.}~\bibnamefont {Liang}},\ and\ \bibinfo {editor} {\bibfnamefont {J.~W.}\ \bibnamefont {Vaughan}}}\ (\bibinfo  {publisher} {Curran Associates, Inc.},\ \bibinfo {year} {2021})\ pp.\ \bibinfo {pages} {2187--2200}\BibitemShut {NoStop}%
\bibitem [{\citenamefont {Ferran~Pousa}\ \emph {et~al.}(2023)\citenamefont {Ferran~Pousa}, \citenamefont {Jalas}, \citenamefont {Kirchen}, \citenamefont {Martinez de~la Ossa}, \citenamefont {Th\'evenet}, \citenamefont {Hudson}, \citenamefont {Larson}, \citenamefont {Huebl}, \citenamefont {Vay},\ and\ \citenamefont {Lehe}}]{Optimas}%
  \BibitemOpen
  \bibfield  {author} {\bibinfo {author} {\bibfnamefont {A.}~\bibnamefont {Ferran~Pousa}}, \bibinfo {author} {\bibfnamefont {S.}~\bibnamefont {Jalas}}, \bibinfo {author} {\bibfnamefont {M.}~\bibnamefont {Kirchen}}, \bibinfo {author} {\bibfnamefont {A.}~\bibnamefont {Martinez de~la Ossa}}, \bibinfo {author} {\bibfnamefont {M.}~\bibnamefont {Th\'evenet}}, \bibinfo {author} {\bibfnamefont {S.}~\bibnamefont {Hudson}}, \bibinfo {author} {\bibfnamefont {J.}~\bibnamefont {Larson}}, \bibinfo {author} {\bibfnamefont {A.}~\bibnamefont {Huebl}}, \bibinfo {author} {\bibfnamefont {J.-L.}\ \bibnamefont {Vay}},\ and\ \bibinfo {author} {\bibfnamefont {R.}~\bibnamefont {Lehe}},\ }\bibfield  {title} {\bibinfo {title} {{Bayesian optimization of laser-plasma accelerators assisted by reduced physical models}},\ }\href {https://doi.org/10.1103/PhysRevAccelBeams.26.084601} {\bibfield  {journal} {\bibinfo  {journal} {Phys. Rev. Accel. Beams}\ }\textbf {\bibinfo {volume} {26}},\ \bibinfo {pages} {084601} (\bibinfo {year} {2023})}\BibitemShut {NoStop}%
\bibitem [{\citenamefont {Hudson}\ \emph {et~al.}(2022)\citenamefont {Hudson}, \citenamefont {Larson}, \citenamefont {Navarro},\ and\ \citenamefont {Wild}}]{LibEnsemble}%
  \BibitemOpen
  \bibfield  {author} {\bibinfo {author} {\bibfnamefont {S.}~\bibnamefont {Hudson}}, \bibinfo {author} {\bibfnamefont {J.}~\bibnamefont {Larson}}, \bibinfo {author} {\bibfnamefont {J.-L.}\ \bibnamefont {Navarro}},\ and\ \bibinfo {author} {\bibfnamefont {S.~M.}\ \bibnamefont {Wild}},\ }\bibfield  {title} {\bibinfo {title} {{libEnsemble}: A library to coordinate the concurrent evaluation of dynamic ensembles of calculations},\ }\href {https://doi.org/10.1109/tpds.2021.3082815} {\bibfield  {journal} {\bibinfo  {journal} {{IEEE} Transactions on Parallel and Distributed Systems}\ }\textbf {\bibinfo {volume} {33}},\ \bibinfo {pages} {977} (\bibinfo {year} {2022})}\BibitemShut {NoStop}%
\bibitem [{\citenamefont {Schulz}\ \emph {et~al.}(2015)\citenamefont {Schulz}, \citenamefont {Grgura{\v s}}, \citenamefont {Behrens}, \citenamefont {Bromberger}, \citenamefont {Costello}, \citenamefont {Czwalinna}, \citenamefont {Felber}, \citenamefont {Hoffmann}, \citenamefont {Ilchen}, \citenamefont {Liu}, \citenamefont {Mazza}, \citenamefont {Meyer}, \citenamefont {Pfeiffer}, \citenamefont {Pr{e}dki}, \citenamefont {Schefer}, \citenamefont {Schmidt}, \citenamefont {Wegner}, \citenamefont {Schlarb},\ and\ \citenamefont {Cavalieri}}]{Schulz2015Timing}%
  \BibitemOpen
  \bibfield  {author} {\bibinfo {author} {\bibfnamefont {S.}~\bibnamefont {Schulz}}, \bibinfo {author} {\bibfnamefont {I.}~\bibnamefont {Grgura{\v s}}}, \bibinfo {author} {\bibfnamefont {C.}~\bibnamefont {Behrens}}, \bibinfo {author} {\bibfnamefont {H.}~\bibnamefont {Bromberger}}, \bibinfo {author} {\bibfnamefont {J.~T.}\ \bibnamefont {Costello}}, \bibinfo {author} {\bibfnamefont {M.~K.}\ \bibnamefont {Czwalinna}}, \bibinfo {author} {\bibfnamefont {M.}~\bibnamefont {Felber}}, \bibinfo {author} {\bibfnamefont {M.~C.}\ \bibnamefont {Hoffmann}}, \bibinfo {author} {\bibfnamefont {M.}~\bibnamefont {Ilchen}}, \bibinfo {author} {\bibfnamefont {H.~Y.}\ \bibnamefont {Liu}}, \bibinfo {author} {\bibfnamefont {T.}~\bibnamefont {Mazza}}, \bibinfo {author} {\bibfnamefont {M.}~\bibnamefont {Meyer}}, \bibinfo {author} {\bibfnamefont {S.}~\bibnamefont {Pfeiffer}}, \bibinfo {author} {\bibfnamefont {P.}~\bibnamefont {Pr{e}dki}}, \bibinfo {author} {\bibfnamefont {S.}~\bibnamefont {Schefer}}, \bibinfo {author} {\bibfnamefont {C.}~\bibnamefont {Schmidt}}, \bibinfo {author} {\bibfnamefont {U.}~\bibnamefont {Wegner}}, \bibinfo {author} {\bibfnamefont {H.}~\bibnamefont {Schlarb}},\ and\ \bibinfo {author} {\bibfnamefont {A.~L.}\ \bibnamefont {Cavalieri}},\ }\bibfield  {title} {\bibinfo {title} {Femtosecond all-optical synchronization of an x-ray free-electron laser},\ }\href@noop {} {\bibfield  {journal} {\bibinfo  {journal} {Nat. Commun.}\ }\textbf {\bibinfo {volume} {6}},\ \bibinfo {pages} {5938} (\bibinfo {year} {2015})}\BibitemShut {NoStop}%
\bibitem [{\citenamefont {Shalloo}\ \emph {et~al.}(2015)\citenamefont {Shalloo}, \citenamefont {Arran}, \citenamefont {Cheung}, \citenamefont {Corner}, \citenamefont {Holloway}, \citenamefont {Walczak}, \citenamefont {Hooker}, \citenamefont {Booth}, \citenamefont {Chekhlov}, \citenamefont {Gregory} \emph {et~al.}}]{shalloo2015synchronisation}%
  \BibitemOpen
  \bibfield  {author} {\bibinfo {author} {\bibfnamefont {R.}~\bibnamefont {Shalloo}}, \bibinfo {author} {\bibfnamefont {C.}~\bibnamefont {Arran}}, \bibinfo {author} {\bibfnamefont {G.}~\bibnamefont {Cheung}}, \bibinfo {author} {\bibfnamefont {L.}~\bibnamefont {Corner}}, \bibinfo {author} {\bibfnamefont {J.}~\bibnamefont {Holloway}}, \bibinfo {author} {\bibfnamefont {R.}~\bibnamefont {Walczak}}, \bibinfo {author} {\bibfnamefont {S.}~\bibnamefont {Hooker}}, \bibinfo {author} {\bibfnamefont {N.}~\bibnamefont {Booth}}, \bibinfo {author} {\bibfnamefont {O.}~\bibnamefont {Chekhlov}}, \bibinfo {author} {\bibfnamefont {C.}~\bibnamefont {Gregory}}, \emph {et~al.},\ }\bibfield  {title} {\bibinfo {title} {{Measurement of femtosecond-scale drift and jitter of the delay between the North and South Beams of Gemini}},\ }\href@noop {} {\bibfield  {journal} {\bibinfo  {journal} {Central Laser Facility Annual Report}\ }\textbf {\bibinfo {volume} {36}} (\bibinfo {year} {2015})}\BibitemShut {NoStop}%
\bibitem [{\citenamefont {Christie}(2024)}]{Christie2024}%
  \BibitemOpen
  \bibfield  {author} {\bibinfo {author} {\bibfnamefont {J.}~\bibnamefont {Christie}},\ }\emph {\bibinfo {title} {Femtosecond Synchronisation for Externally-Injected Laser Wakefield Acceleration at CLARA}},\ \href@noop {} {\bibinfo {type} {{PhD thesis}}},\ \bibinfo  {school} {University of Liverpool} (\bibinfo {year} {2024})\BibitemShut {NoStop}%
\bibitem [{\citenamefont {Winkler}\ \emph {et~al.}(2025)\citenamefont {Winkler}, \citenamefont {Trunk}, \citenamefont {Hübner}, \citenamefont {Ossa}, \citenamefont {Jalas}, \citenamefont {Kirchen}, \citenamefont {Agapov}, \citenamefont {Antipov}, \citenamefont {Brinkmann}, \citenamefont {Eichner}, \citenamefont {Pousa}, \citenamefont {Hülsenbusch}, \citenamefont {Palmer}, \citenamefont {Schnepp}, \citenamefont {Schubert}, \citenamefont {Thévenet}, \citenamefont {Walker}, \citenamefont {Werle}, \citenamefont {Leemans},\ and\ \citenamefont {Maier}}]{Winkler.2025}%
  \BibitemOpen
  \bibfield  {author} {\bibinfo {author} {\bibfnamefont {P.}~\bibnamefont {Winkler}}, \bibinfo {author} {\bibfnamefont {M.}~\bibnamefont {Trunk}}, \bibinfo {author} {\bibfnamefont {L.}~\bibnamefont {Hübner}}, \bibinfo {author} {\bibfnamefont {A.~M. d.~l.}\ \bibnamefont {Ossa}}, \bibinfo {author} {\bibfnamefont {S.}~\bibnamefont {Jalas}}, \bibinfo {author} {\bibfnamefont {M.}~\bibnamefont {Kirchen}}, \bibinfo {author} {\bibfnamefont {I.}~\bibnamefont {Agapov}}, \bibinfo {author} {\bibfnamefont {S.~A.}\ \bibnamefont {Antipov}}, \bibinfo {author} {\bibfnamefont {R.}~\bibnamefont {Brinkmann}}, \bibinfo {author} {\bibfnamefont {T.}~\bibnamefont {Eichner}}, \bibinfo {author} {\bibfnamefont {A.~F.}\ \bibnamefont {Pousa}}, \bibinfo {author} {\bibfnamefont {T.}~\bibnamefont {Hülsenbusch}}, \bibinfo {author} {\bibfnamefont {G.}~\bibnamefont {Palmer}}, \bibinfo {author} {\bibfnamefont {M.}~\bibnamefont {Schnepp}}, \bibinfo {author} {\bibfnamefont {K.}~\bibnamefont {Schubert}}, \bibinfo {author} {\bibfnamefont {M.}~\bibnamefont {Thévenet}}, \bibinfo {author} {\bibfnamefont {P.~A.}\ \bibnamefont {Walker}}, \bibinfo {author} {\bibfnamefont {C.}~\bibnamefont {Werle}}, \bibinfo {author} {\bibfnamefont {W.~P.}\ \bibnamefont {Leemans}},\ and\ \bibinfo {author} {\bibfnamefont {A.~R.}\ \bibnamefont {Maier}},\ }\bibfield  {title} {\bibinfo {title} {{Active energy compression of a laser-plasma electron beam}},\ }\href {https://doi.org/10.1038/s41586-025-08772-y} {\bibfield  {journal} {\bibinfo  {journal} {Nature}\ ,\ \bibinfo {pages} {1}} (\bibinfo {year} {2025})}\BibitemShut {NoStop}%
\bibitem [{\citenamefont {Di~Mitri}(2015)}]{DiMitri2015}%
  \BibitemOpen
  \bibfield  {author} {\bibinfo {author} {\bibfnamefont {S.}~\bibnamefont {Di~Mitri}},\ }\bibfield  {title} {\bibinfo {title} {{On the Importance of Electron Beam Brightness in High Gain Free Electron Lasers}},\ }\href {https://doi.org/10.3390/photonics2020317} {\bibfield  {journal} {\bibinfo  {journal} {Photonics}\ }\textbf {\bibinfo {volume} {2}},\ \bibinfo {pages} {317} (\bibinfo {year} {2015})}\BibitemShut {NoStop}%
\bibitem [{\citenamefont {Huang}\ and\ \citenamefont {Kim}(2007)}]{Huang2007}%
  \BibitemOpen
  \bibfield  {author} {\bibinfo {author} {\bibfnamefont {Z.}~\bibnamefont {Huang}}\ and\ \bibinfo {author} {\bibfnamefont {K.-J.}\ \bibnamefont {Kim}},\ }\bibfield  {title} {\bibinfo {title} {Review of x-ray free-electron laser theory},\ }\href {https://doi.org/10.1103/PhysRevSTAB.10.034801} {\bibfield  {journal} {\bibinfo  {journal} {Phys. Rev. ST Accel. Beams}\ }\textbf {\bibinfo {volume} {10}},\ \bibinfo {pages} {034801} (\bibinfo {year} {2007})}\BibitemShut {NoStop}%
\bibitem [{\citenamefont {Tiedtke}\ \emph {et~al.}(2009)\citenamefont {Tiedtke}, \citenamefont {Azima}, \citenamefont {von Bargen}, \citenamefont {Bittner}, \citenamefont {Bonfigt}, \citenamefont {Düsterer}, \citenamefont {Faatz}, \citenamefont {Frühling}, \citenamefont {Gensch}, \citenamefont {Gerth}, \citenamefont {Guerassimova}, \citenamefont {Hahn}, \citenamefont {Hans}, \citenamefont {Hesse}, \citenamefont {Honkavaar}, \citenamefont {Jastrow}, \citenamefont {Juranic}, \citenamefont {Kapitzki}, \citenamefont {Keitel}, \citenamefont {Kracht}, \citenamefont {Kuhlmann}, \citenamefont {Li}, \citenamefont {Martins}, \citenamefont {Núñez}, \citenamefont {Plönjes}, \citenamefont {Redlin}, \citenamefont {Saldin}, \citenamefont {Schneidmiller}, \citenamefont {Schneider}, \citenamefont {Schreiber}, \citenamefont {Stojanovic}, \citenamefont {Tavella}, \citenamefont {Toleikis}, \citenamefont {Treusch}, \citenamefont {Weigelt}, \citenamefont {Wellhöfer}, \citenamefont {Wabnitz}, \citenamefont {Yurkov},\ and\ \citenamefont {Feldhaus}}]{FLASHTiedtke2009}%
  \BibitemOpen
  \bibfield  {author} {\bibinfo {author} {\bibfnamefont {K.}~\bibnamefont {Tiedtke}}, \bibinfo {author} {\bibfnamefont {A.}~\bibnamefont {Azima}}, \bibinfo {author} {\bibfnamefont {N.}~\bibnamefont {von Bargen}}, \bibinfo {author} {\bibfnamefont {L.}~\bibnamefont {Bittner}}, \bibinfo {author} {\bibfnamefont {S.}~\bibnamefont {Bonfigt}}, \bibinfo {author} {\bibfnamefont {S.}~\bibnamefont {Düsterer}}, \bibinfo {author} {\bibfnamefont {B.}~\bibnamefont {Faatz}}, \bibinfo {author} {\bibfnamefont {U.}~\bibnamefont {Frühling}}, \bibinfo {author} {\bibfnamefont {M.}~\bibnamefont {Gensch}}, \bibinfo {author} {\bibfnamefont {C.}~\bibnamefont {Gerth}}, \bibinfo {author} {\bibfnamefont {N.}~\bibnamefont {Guerassimova}}, \bibinfo {author} {\bibfnamefont {U.}~\bibnamefont {Hahn}}, \bibinfo {author} {\bibfnamefont {T.}~\bibnamefont {Hans}}, \bibinfo {author} {\bibfnamefont {M.}~\bibnamefont {Hesse}}, \bibinfo {author} {\bibfnamefont {K.}~\bibnamefont {Honkavaar}}, \bibinfo {author} {\bibfnamefont {U.}~\bibnamefont {Jastrow}}, \bibinfo {author} {\bibfnamefont {P.}~\bibnamefont {Juranic}}, \bibinfo {author} {\bibfnamefont {S.}~\bibnamefont {Kapitzki}}, \bibinfo {author} {\bibfnamefont {B.}~\bibnamefont {Keitel}}, \bibinfo {author} {\bibfnamefont {T.}~\bibnamefont {Kracht}}, \bibinfo {author} {\bibfnamefont {M.}~\bibnamefont {Kuhlmann}}, \bibinfo {author} {\bibfnamefont {W.~B.}\ \bibnamefont {Li}}, \bibinfo {author} {\bibfnamefont {M.}~\bibnamefont {Martins}}, \bibinfo {author} {\bibfnamefont {T.}~\bibnamefont {Núñez}}, \bibinfo {author} {\bibfnamefont {E.}~\bibnamefont {Plönjes}}, \bibinfo {author} {\bibfnamefont {H.}~\bibnamefont {Redlin}}, \bibinfo {author} {\bibfnamefont {E.~L.}\ \bibnamefont {Saldin}}, \bibinfo {author} {\bibfnamefont {E.~A.}\ \bibnamefont {Schneidmiller}}, \bibinfo {author} {\bibfnamefont {J.~R.}\ \bibnamefont {Schneider}}, \bibinfo {author} {\bibfnamefont {S.}~\bibnamefont {Schreiber}}, \bibinfo {author} {\bibfnamefont {N.}~\bibnamefont {Stojanovic}}, \bibinfo {author} {\bibfnamefont {F.}~\bibnamefont
  {Tavella}}, \bibinfo {author} {\bibfnamefont {S.}~\bibnamefont {Toleikis}}, \bibinfo {author} {\bibfnamefont {R.}~\bibnamefont {Treusch}}, \bibinfo {author} {\bibfnamefont {H.}~\bibnamefont {Weigelt}}, \bibinfo {author} {\bibfnamefont {M.}~\bibnamefont {Wellhöfer}}, \bibinfo {author} {\bibfnamefont {H.}~\bibnamefont {Wabnitz}}, \bibinfo {author} {\bibfnamefont {M.~V.}\ \bibnamefont {Yurkov}},\ and\ \bibinfo {author} {\bibfnamefont {J.}~\bibnamefont {Feldhaus}},\ }\bibfield  {title} {\bibinfo {title} {{The soft x-ray free-electron laser FLASH at DESY: beamlines, diagnostics and end-stations}},\ }\href {https://doi.org/10.1088/1367-2630/11/2/023029} {\bibfield  {journal} {\bibinfo  {journal} {New Journal of Physics}\ }\textbf {\bibinfo {volume} {11}},\ \bibinfo {pages} {023029} (\bibinfo {year} {2009})}\BibitemShut {NoStop}%
\bibitem [{\citenamefont {Bourgeois}\ \emph {et~al.}(2013)\citenamefont {Bourgeois}, \citenamefont {Cowley},\ and\ \citenamefont {Hooker}}]{Bourgeois2013}%
  \BibitemOpen
  \bibfield  {author} {\bibinfo {author} {\bibfnamefont {N.}~\bibnamefont {Bourgeois}}, \bibinfo {author} {\bibfnamefont {J.}~\bibnamefont {Cowley}},\ and\ \bibinfo {author} {\bibfnamefont {S.~M.}\ \bibnamefont {Hooker}},\ }\bibfield  {title} {\bibinfo {title} {{Two-Pulse Ionization Injection into Quasilinear Laser Wakefields}},\ }\href {https://doi.org/10.1103/PhysRevLett.111.155004} {\bibfield  {journal} {\bibinfo  {journal} {Phys. Rev. Lett.}\ }\textbf {\bibinfo {volume} {111}},\ \bibinfo {pages} {155004} (\bibinfo {year} {2013})}\BibitemShut {NoStop}%
\bibitem [{\citenamefont {Yu}\ \emph {et~al.}(2014)\citenamefont {Yu}, \citenamefont {Esarey}, \citenamefont {Schroeder}, \citenamefont {Vay}, \citenamefont {Benedetti}, \citenamefont {Geddes}, \citenamefont {Chen},\ and\ \citenamefont {Leemans}}]{Yu2014}%
  \BibitemOpen
  \bibfield  {author} {\bibinfo {author} {\bibfnamefont {L.-L.}\ \bibnamefont {Yu}}, \bibinfo {author} {\bibfnamefont {E.}~\bibnamefont {Esarey}}, \bibinfo {author} {\bibfnamefont {C.~B.}\ \bibnamefont {Schroeder}}, \bibinfo {author} {\bibfnamefont {J.-L.}\ \bibnamefont {Vay}}, \bibinfo {author} {\bibfnamefont {C.}~\bibnamefont {Benedetti}}, \bibinfo {author} {\bibfnamefont {C.~G.~R.}\ \bibnamefont {Geddes}}, \bibinfo {author} {\bibfnamefont {M.}~\bibnamefont {Chen}},\ and\ \bibinfo {author} {\bibfnamefont {W.~P.}\ \bibnamefont {Leemans}},\ }\bibfield  {title} {\bibinfo {title} {{Two-Color Laser-Ionization Injection}},\ }\href {https://doi.org/10.1103/PhysRevLett.112.125001} {\bibfield  {journal} {\bibinfo  {journal} {Phys. Rev. Lett.}\ }\textbf {\bibinfo {volume} {112}},\ \bibinfo {pages} {125001} (\bibinfo {year} {2014})}\BibitemShut {NoStop}%
\bibitem [{\citenamefont {Hooker}\ \emph {et~al.}(2014)\citenamefont {Hooker}, \citenamefont {Bartolini}, \citenamefont {Mangles}, \citenamefont {Tünnermann}, \citenamefont {Corner}, \citenamefont {Limpert}, \citenamefont {Seryi},\ and\ \citenamefont {Walczak}}]{Hooker2014MPLWFA}%
  \BibitemOpen
  \bibfield  {author} {\bibinfo {author} {\bibfnamefont {S.~M.}\ \bibnamefont {Hooker}}, \bibinfo {author} {\bibfnamefont {R.}~\bibnamefont {Bartolini}}, \bibinfo {author} {\bibfnamefont {S.~P.~D.}\ \bibnamefont {Mangles}}, \bibinfo {author} {\bibfnamefont {A.}~\bibnamefont {Tünnermann}}, \bibinfo {author} {\bibfnamefont {L.}~\bibnamefont {Corner}}, \bibinfo {author} {\bibfnamefont {J.}~\bibnamefont {Limpert}}, \bibinfo {author} {\bibfnamefont {A.}~\bibnamefont {Seryi}},\ and\ \bibinfo {author} {\bibfnamefont {R.}~\bibnamefont {Walczak}},\ }\bibfield  {title} {\bibinfo {title} {Multi-pulse laser wakefield acceleration: a new route to efficient, high-repetition-rate plasma accelerators and high flux radiation sources},\ }\href {https://doi.org/10.1088/0953-4075/47/23/234003} {\bibfield  {journal} {\bibinfo  {journal} {J. Phys. B: At. Mol. Opt. Phys.}\ }\textbf {\bibinfo {volume} {47}},\ \bibinfo {pages} {234003} (\bibinfo {year} {2014})}\BibitemShut {NoStop}%
\bibitem [{\citenamefont {Jakobsson}\ \emph {et~al.}(2021)\citenamefont {Jakobsson}, \citenamefont {Hooker},\ and\ \citenamefont {Walczak}}]{Jakobsson2021}%
  \BibitemOpen
  \bibfield  {author} {\bibinfo {author} {\bibfnamefont {O.}~\bibnamefont {Jakobsson}}, \bibinfo {author} {\bibfnamefont {S.~M.}\ \bibnamefont {Hooker}},\ and\ \bibinfo {author} {\bibfnamefont {R.}~\bibnamefont {Walczak}},\ }\bibfield  {title} {\bibinfo {title} {{GeV-Scale Accelerators Driven by Plasma-Modulated Pulses from Kilohertz Lasers}},\ }\href {https://doi.org/10.1103/PhysRevLett.127.184801} {\bibfield  {journal} {\bibinfo  {journal} {Phys. Rev. Lett.}\ }\textbf {\bibinfo {volume} {127}},\ \bibinfo {pages} {184801} (\bibinfo {year} {2021})}\BibitemShut {NoStop}%
\bibitem [{\citenamefont {Diederichs}\ \emph {et~al.}(2022)\citenamefont {Diederichs}, \citenamefont {Benedetti}, \citenamefont {Huebl}, \citenamefont {Lehe}, \citenamefont {Myers}, \citenamefont {Sinn}, \citenamefont {Vay}, \citenamefont {Zhang},\ and\ \citenamefont {Thévenet}}]{Diederichs2022HiPACE++}%
  \BibitemOpen
  \bibfield  {author} {\bibinfo {author} {\bibfnamefont {S.}~\bibnamefont {Diederichs}}, \bibinfo {author} {\bibfnamefont {C.}~\bibnamefont {Benedetti}}, \bibinfo {author} {\bibfnamefont {A.}~\bibnamefont {Huebl}}, \bibinfo {author} {\bibfnamefont {R.}~\bibnamefont {Lehe}}, \bibinfo {author} {\bibfnamefont {A.}~\bibnamefont {Myers}}, \bibinfo {author} {\bibfnamefont {A.}~\bibnamefont {Sinn}}, \bibinfo {author} {\bibfnamefont {J.-L.}\ \bibnamefont {Vay}}, \bibinfo {author} {\bibfnamefont {W.}~\bibnamefont {Zhang}},\ and\ \bibinfo {author} {\bibfnamefont {M.}~\bibnamefont {Thévenet}},\ }\bibfield  {title} {\bibinfo {title} {{HiPACE++: A portable, 3D quasi-static particle-in-cell code}},\ }\href {https://doi.org/https://doi.org/10.1016/j.cpc.2022.108421} {\bibfield  {journal} {\bibinfo  {journal} {Computer Physics Communications}\ }\textbf {\bibinfo {volume} {278}},\ \bibinfo {pages} {108421} (\bibinfo {year} {2022})}\BibitemShut {NoStop}%
\end{thebibliography}%

\end{document}